\documentclass[11pt,prd,onecolumn,amsmath,amssymb,aps,floats,floatfix,nofootinbib]{revtex4-2}

\usepackage[colorlinks=true,urlcolor=blue,anchorcolor=blue,citecolor=blue,filecolor=blue,linkcolor=blue,menucolor=blue,linktocpage=true]{hyperref} 

\usepackage[inline]{enumitem}
\usepackage[multidot]{grffile}  
\usepackage{dcolumn}
\usepackage{bm}
\usepackage{amsmath}
\usepackage{amsfonts}
\usepackage{amssymb}
\usepackage{color}
\usepackage[table]{xcolor}
\usepackage{float}
\usepackage{latexsym}
\usepackage{slashed} 
\usepackage{pstricks}
\usepackage{indentfirst}
\usepackage{mathrsfs}
\usepackage{multirow}
\usepackage{epsfig,psfrag}
\usepackage{graphicx}
\usepackage{enumitem}
\usepackage{geometry,amssymb,yfonts}
\usepackage{yhmath}
\usepackage{anysize}
\usepackage{subfigure}
\usepackage{mathtools}
\usepackage{setspace} 
\usepackage[utf8]{inputenc} 
\usepackage[scientific-notation=true]{siunitx} 

\graphicspath{{fig/}}

\allowdisplaybreaks 
\definecolor{myred}{rgb}{.75,0,0}

\begin{document}

\title{Gravitational waves from graviton bremsstrahlung in scalar leptoquark decays}

\author{Qian-Jiu Wang}
\author{Hua Tong}
\author{Zhao-Huan Yu}\email[Corresponding author. ]{yuzhaoh5@mail.sysu.edu.cn}
\affiliation{School of Physics, Sun Yat-Sen University, Guangzhou 510275, China}

\begin{abstract}
We study the stochastic gravitational wave background originated from graviton bremsstrahlung in decays of scalar leptoquarks, which are colored scalar bosons simultaneously coupling to a quark and a lepton. We take the scalar leptoquarks in the $\mathrm{SU}(5)$ grand unified theory as a concrete example. Stringent experimental bounds on proton decay force these particles to be superheavy, which in turn renders their graviton bremsstrahlung, induced by quantum gravity effects, less suppressed. By solving the relevant Boltzmann equation, we trace the evolution of the scalar leptoquark number density in the early Universe and use it to compute the resulting gravitational wave spectrum. We find that if the reheating temperature is sufficiently high, the signal would be significant and could be probed by high-frequency gravitational wave detectors employing resonant cavity techniques.
\end{abstract}

\maketitle
\tableofcontents

\clearpage

\section{Introduction}

The discovery of gravitational waves (GWs) by the Laser Interferometer Gravitational Observatory (LIGO) in 2015 \cite{LIGOScientific:2016aoc} opened a unique window on to the Universe.
GWs now serve as a new cosmic messenger for exploring new physics beyond the standard model (SM).
Unlike electromagnetic radiation, which was opaque in the early Universe until photon decoupling, GWs propagate freely through space, carrying pristine information from the earliest epochs directly to modern observatories.

New physics theories typically predict heavy particles that were produced in the early Universe.
Their decays and scattering are expected to be accompanied by graviton bremsstrahlung via quantum gravitational processes \cite{Weinberg:1965nx,Nakayama:2018ptw}, thereby inducing a stochastic GW background (SGWB) that is potentially detectable.
Relevant processes include inflaton decays  \cite{Nakayama:2018ptw,Huang:2019lgd,Ghoshal:2022kqp,Barman:2023ymn,Tokareva:2023mrt,Jiang:2024akb,Cheng:2025gyh,Das:2025cqs}, inflaton scattering \cite{Xu:2024fjl,Xu:2025wjq}, decays of right-handed neutrinos during leptogenesis \cite{Datta:2024tne,Choi:2025hqt,Murayama:2025thw}, freeze-in production of dark matter \cite{Konar:2025iuk,Wang:2025lmf,Konar:2025gvh}, particle scattering in the thermal plasma \cite{Ghiglieri:2015nfa,Ghiglieri:2020mhm,Ringwald:2020ist,Drewes:2023oxg,Ghiglieri:2024ghm,Bernal:2025lxp,Montefalcone:2025gxx,Chen:2025try}, as well as decays of superheavy particles from preheating \cite{Hu:2024awd,Kanemura:2025rct}, from evaporation of primordial black holes \cite{Choi:2024acs}, from the type-II seesaw model \cite{Wang:2025mtq}, and from axion models \cite{Wang:2026qsg}.

The branching ratio for particle decays that emit a bremsstrahlung graviton is suppressed by the square of the ratio of the parent particle mass to the Planck mass, within the effective quantum field theory based on Einstein gravity \cite{Nakayama:2018ptw}.
Consequently, to achieve an appreciable SGWB for GW observatories, the parent particle needs to be sufficiently heavy.
For a parent particle with a mass of $10^{12}~\si{GeV}$, the resulting SGWB spectrum is typically predicted to span the frequency range of $10^7$ to $10^{12}~\si{Hz}$ \cite{Konar:2025iuk,Wang:2026qsg}.
Such a prediction overlaps with the $10^{5}\text{--}10^{9}~\si{Hz}$ sensitivity band of proposed high-frequency GW experiments that employ resonant cavity techniques \cite{Herman:2020wao,Berlin:2021txa,Herman:2022fau,Berlin:2023grv,Gatti:2024mde,Cai:2025fpe}.
Therefore, this provides a unique avenue for probing superheavy particles far above the weak scale, a class of particles that is challenging to detect in conventional particle physics experiments.

In this work, we focus on graviton bremsstrahlung in decays of scalar leptoquarks, which are scalar bosons simultaneously coupling to a quark and a lepton \cite{Dorsner:2016wpm}.
They commonly appear in the Higgs sectors of grand unified theories (GUTs) \cite{Georgi:1974sy,Slansky:1981yr}.
Interactions of scalar leptoquarks violate both lepton and baryon numbers, triggering proton decay.
Consequently, stringent experimental limits on proton decay have pushed their lower mass bounds to $\mathcal{O}(10^{11})\text{--}\mathcal{O}(10^{13})~\si{GeV}$ \cite{Dorsner:2012nq}.
The graviton bremsstrahlung from decays of such superheavy leptoquarks thus offers a promising source for an SGWB.

As an illustrative case study, we consider the color-triplet scalar leptoquarks from the $\mathbf{5}$ representation in the $\mathrm{SU}(5)$ GUT.
Their masses are constrained by experimental searches for proton decay.
We then compute the evolution of their number density in the early Universe by solving the Boltzmann equation.
Based on this number density, the spectrum of the SGWB generated via graviton bremsstrahlung in scalar leptoquark decays is estimated.
Finally, we assess the detection prospects for this SGWB at future GW observatories.

The rest of the paper is organized as follows.
In Sec.~\ref{sub:setup}, we describe the $\mathrm{SU}(5)$ GUT framework.
Section~\ref{sub:proton_decay} discusses the constraints imposed by proton decay on the parameter space.
In Sec.~\ref{sub:num_den}, we solve the Boltzmann equation to compute the number density of scalar leptoquarks in the early Universe.
In Sec.~\ref{sub:GW}, the SGWB spectra produced by graviton bremsstrahlung in scalar leptoquark decays are presented.
Section~\ref{sub:Conclusions} gives the conclusions.

\section{The $\mathrm{SU(5)}$ Setup}
\label{sub:setup}

As the first attempt at grand unification, the Georgi-Glashow model \cite{Georgi:1974sy} embeds the SM gauge group $\mathrm{SU(3)_C}\times \mathrm{SU(2)_L}\times \mathrm{U(1)_Y}$ into the $\mathrm{SU(5)}$ group, unifying quarks and leptons within its irreducible representations.
However, the minimal $\mathrm{SU(5)}$ realization has been ruled out due to its incorrect predictions for fermion masses and mixing parameters, as well as its failure to achieve precise gauge coupling unification. 
To rescue the $\mathrm{SU(5)}$ GUT, extended versions incorporating higher-dimensional operators \cite{Ellis:1979fg,Weinberg:1979sa,Shafi:1983gz,Senjanovic:2024uzn} and additional fields \cite{Georgi:1979df,Glashow:1979nm,Dorsner:2005fq,Bajc:2006ia,Dorsner:2007fy,Dorsner:2012nq,FileviezPerez:2016sal,Hagedorn:2016dze,Dorsner:2019vgf,Dorsner:2024jiy,Fang:2024mfn} have been extensively explored.

In the work, our primary focus is on the properties of scalar leptoquarks, rather than on constructing a complete $\mathrm{SU(5)}$ GUT that satisfies all phenomenological requirements.
To proceed, we adopt a simple framework in which 
the masses of SM fermions are generated by Higgs fields in the $\mathbf{5}_\mathrm{H}$ and $\mathbf{45}_\mathrm{H}$ representations, where the subscript $\mathrm{H}$ denotes Higgs representations.
The vacuum expectation value (VEV) of a $\mathbf{24}_\mathrm{H}$ Higgs multiplet at the GUT scale $\sim \mathcal{O}(10^{15})~\si{GeV}$ breaks the $\mathrm{SU(5)}$ gauge symmetry down to $\mathrm{SU(3)_C}\times \mathrm{SU(2)_L}\times \mathrm{U(1)_Y}$, while the VEVs of $\mathbf{5}_\mathrm{H}$ and $\mathbf{45}_\mathrm{H}$ at the electroweak scale further break the gauge symmetry to $\mathrm{SU(3)_C}\times \mathrm{U(1)_{EM}}$.
Three generations of right-handed neutrinos are introduced as three fermionic $\mathrm{SU(5)}$ singlets $\mathbf{1}_{\mathrm{F},i}$ ($i=1,2,3$), and the tiny active neutrino masses are generated via the type-I seesaw mechanism \cite{Minkowski:1977sc,Gell-Mann:1979vob,Yanagida:1979as}.
This constitutes a minimal, renormalizable setup capable of reproducing viable fermion masses and mixing patterns \cite{Georgi:1979df,Dorsner:2012nq}.

The $\mathbf{5}_\mathrm{H}$ Higgs multiplet, with components labeled by an $\mathrm{SU(5)}$ index $\alpha = 1,2,\cdots,5$, decomposes as
\begin{equation}
(\mathbf{5}_\mathrm{H})^\alpha = (\Delta_1, \Delta_2, \Delta_3, H_1, H_2) = \Delta (\mathbf{3},\mathbf{1},-1/3) \oplus H (\mathbf{1},\mathbf{2},1/2).
\end{equation}
In the last expression, the representations of the $\mathrm{SU(3)_C}\times \mathrm{SU(2)_L}\times \mathrm{U(1)_Y}$ gauge group are indicated.
The $\mathrm{SU(3)_C}$ triplet $\Delta$ contains three scalar leptoquarks $\Delta_a$, where $a=1,2,3$ is a color index.
In the following, we will investigate the SGWB generated by graviton bremsstrahlung in $\Delta_a$ decays as a concrete phenomenological example.
The $\mathrm{SU(2)_L}$ doublet $H$ acquires a VEV $\langle (\mathbf{5}_\mathrm{H})^5  \rangle = \langle H_2 \rangle = v_5/\sqrt{2}$, contributing to the spontaneous breaking of the electroweak gauge symmetry.
An additional contribution arises from the $\mathbf{45}_\mathrm{H}$ Higgs multiplet, which can be represented by a rank-3 $\mathrm{SU(5)}$ tensor $(\mathbf{45}_\mathrm{H})^{\alpha\beta}_\gamma$ satisfying the antisymmetry and traceless conditions \cite{Georgi:1979df,Dorsner:2007fy}
\begin{equation}
(\mathbf{45}_\mathrm{H})^{\alpha\beta}_\gamma = - (\mathbf{45}_\mathrm{H})^{\beta\alpha}_\gamma,\quad
\sum_{\alpha=1}^5 (\mathbf{45}_\mathrm{H})^{\alpha\beta}_\alpha = 0,\quad
\alpha,\beta,\gamma = 1,2,3,4,5.
\end{equation}
An appropriate VEV configuration is given by
\begin{equation}
\langle (\mathbf{45}_\mathrm{H})^{\alpha 5}_{\beta} \rangle = \frac{v_{45}}{\sqrt{2}} (\delta^\alpha_\beta - 4\delta^{\alpha 4}\delta_{\beta 4}),\quad \alpha,\beta=1,2,3,4,
\label{equ:the vev of Higgs in the 45-dimentional representation}
\end{equation}
yielding the relations $\langle (\mathbf{45}_\mathrm{H})^{15}_{1} \rangle = \langle (\mathbf{45}_\mathrm{H})^{25}_{2} \rangle = \langle (\mathbf{45}_\mathrm{H})^{35}_{3} \rangle = v_{45}/\sqrt{2} = - \langle (\mathbf{45}_\mathrm{H})^{45}_{4} \rangle/3$.

The covariant derivatives of the $\mathbf{5}_\mathrm{H}$ and $\mathbf{45}_\mathrm{H}$ Higgs multiplets can be expressed as
\begin{eqnarray}
(D_{\mu} \mathbf{5}_{\mathrm{H}})^{\alpha} &=& \partial_{\mu}
(\mathbf{5}_{\mathrm{H}})^{\alpha} - \mathrm{i} g_5 (A_{\mu})^{\alpha}_{\beta}
(\mathbf{5}_{\mathrm{H}})^{\beta},
\\
 (D_{\mu} \mathbf{45}_{\mathrm{H}})^{\alpha \beta}_{\gamma} & = &
  \partial_{\mu} (\mathbf{45}_{\mathrm{H}})^{\alpha \beta}_{\gamma} - \mathrm{i}
  g_5 (A_{\mu})^{\alpha}_{\delta} (\mathbf{45}_{\mathrm{H}})^{\delta
  \beta}_{\gamma} - \mathrm{i} g_5 (A_{\mu})^{\beta}_{\delta}
  (\mathbf{45}_{\mathrm{H}})^{\alpha \delta}_{\gamma} + \mathrm{i} g_5
  (A_{\mu})^{\delta}_{\gamma} (\mathbf{45}_{\mathrm{H}})^{\alpha
  \beta}_{\delta},
\end{eqnarray}
where $g_5$ denotes the $\mathrm{SU(5)}$ gauge coupling, and $(A_{\mu})^{\alpha}_{\beta}$ are the $\mathrm{SU(5)}$ gauge fields in the $\mathbf{24}_\mathrm{V}$ representation.
In particular, the components $(A_{\mu})^4_5 = {W_{\mu}^+}/{\sqrt{2}}$ and $(A_{\mu})^5_4 =
{W_{\mu}^-}/{\sqrt{2}}$ correspond to the $W^\pm$ gauge fields.
After electroweak symmetry breaking, the VEVs of $\mathbf{5}_\mathrm{H}$ and $\mathbf{45}_\mathrm{H}$ contribute to the $W^\pm$ boson mass.
The corresponding mass term in the Lagrangian is
\begin{equation}
\mathcal{L}\supset \left(\frac{1}{4}g_5^2 v_5^2 + 6g_5^2 v_{45}^2\right) W^{\mu,+}W_\mu^{-} = m_{W}^2 W^{\mu,+}W_\mu^{-} .
\end{equation}
The $W^\pm$ boson mass is then given by
\begin{equation}
m_{W} = \frac{1}{2}g_5 v,
\end{equation}
with the effective electroweak VEV defined as $v \equiv \sqrt{v_5^2 + 24v_{45}^2} = 246.22~\si{GeV}$.

The SM fermions are embedded in three families of fermionic $\mathrm{SU(5)}$ multiplets, $\mathbf{5}_{\mathrm{F},i}$ and $\mathbf{10}_{\mathrm{F},i}$,  with the index $i=1,2,3$ denoting the generation.
The $\mathbf{5}_{\mathrm{F},i}$ multiplet decomposes under the SM gauge group as
\begin{equation}
(\mathbf{5}_{\mathrm{F},i})^\alpha = \left( d_{1i}, d_{2i},d_{3i},\ell_i^\mathrm{C},-\nu_{i}^\mathrm{C} \right)_\mathrm{R}
= d_{i\mathrm{R}} (\mathbf{3},\mathbf{1},-1/3) \oplus (L_{i\mathrm{L}})^\mathrm{C} (\mathbf{1},\mathbf{2},1/2),
\end{equation}
where the subscripts $\mathrm{L}$ and $\mathrm{R}$ denote left- and right-handed projections, respectively, and the superscript $\mathrm{C}$ indicates the charge conjugate.
Note that for a four-component spinor $\psi$, we have $(\psi^\mathrm{C})_\mathrm{L/R} = (\psi_\mathrm{R/L})^\mathrm{C}$.
The $\mathbf{10}_{\mathrm{F},i}$ multiplet decomposes as
\begin{eqnarray}
(\mathbf{10}_{\mathrm{F},i})^{\alpha\beta} &=& \frac{1}{\sqrt{2}} \begin{pmatrix}
 0 &  u_{3i}^\mathrm{C} & -u_{2i}^\mathrm{C} & -u_{1i} & -d_{1i} \\
-u_{3i}^\mathrm{C} & 0 & u_{1i}^\mathrm{C} & -u_{2i} & -d_{2i} \\
 u_{2i}^\mathrm{C} &  -u_{1i}^\mathrm{C} & 0 & -u_{3i} & -d_{3i} \\
 u_{1i} & u_{2i} & u_{3i} & 0 & -\ell_i^\mathrm{C} \\
 d_{1i} & d_{2i} & d_{3i} & \ell_i^\mathrm{C} & 0
 \end{pmatrix}_\mathrm{L}
\nonumber\\
&=& (u_{i\mathrm{R}})^\mathrm{C} (\bar{\mathbf{3}},\mathbf{1},-2/3) \oplus Q_{i\mathrm{L}} (\mathbf{3},\mathbf{2},1/6) \oplus (\ell_{i\mathrm{R}})^\mathrm{C} (\mathbf{1},\mathbf{1},1).
\end{eqnarray}
The right-handed neutrinos correspond to $\mathbf{1}_{\mathrm{F}, i} = N_{i \mathrm{R}} (\mathbf{1}, \mathbf{1}, 0)$.

The renormalizable Yukawa couplings among $\mathbf{5}_\mathrm{H}$, $\mathbf{45}_\mathrm{H}$, $\mathbf{5}_{\mathrm{F},i}$, $\mathbf{10}_{\mathrm{F},i}$, and $\mathbf{1}_{\mathrm{F}, i}$, together with the Majorana mass terms for $\mathbf{1}_{\mathrm{F}, i}$, are given by \cite{Dorsner:2007fy,Dorsner:2012nq}
\begin{eqnarray}
\mathcal{L} &\supset& -Y_{ij}^{10} \varepsilon_{\alpha\beta\gamma\delta\epsilon} (\overline{\mathbf{10}^\mathrm{C}_{\mathrm{F},i}})^{\alpha\beta} (\mathbf{10}_{\mathrm{F},j})^{\gamma\delta}(\mathbf{5}_\mathrm{H})^\epsilon + Y_{ij}^{5} (\overline{\mathbf{5}_{\mathrm{F},j}})_\alpha (\mathbf{10}_{\mathrm{F},i})^{\alpha\beta} (\mathbf{5}_\mathrm{H}^*)_\beta
\nonumber\\
&& + Y_{ij}^{45} (\overline{\mathbf{5}_{\mathrm{F},j}})_\gamma (\mathbf{10}_{\mathrm{F},i})^{\alpha\beta} (\mathbf{45}_\mathrm{H}^*)_{\alpha\beta}^\gamma
+ \tilde{Y}_{i j}^{45} \varepsilon_{\alpha \beta \gamma \delta \epsilon}
(\overline{\mathbf{10}^{\mathrm{C}}_{\mathrm{F}, i}})^{\alpha \beta}
(\mathbf{10}_{\mathrm{F}, j})^{\zeta \gamma}
(\mathbf{45}_{\mathrm{H}})^{\delta \varepsilon}_{\zeta}
\nonumber\\
&& + Y_{i j}^1 (\overline{\mathbf{5}_{\mathrm{F}, i}})_{\alpha}
\mathbf{1}^{\mathrm{C}}_{\mathrm{F}, j} (\mathbf{5}_{\mathrm{H}})^{\alpha}
- M_{N, i j} \overline{\mathbf{1}^{\mathrm{C}}_{\mathrm{F}, i}}
\mathbf{1}_{\mathrm{F}, j}
+ \mathrm{H.c.}
\label{eq:L:Yukawa_Mass}
\end{eqnarray}
After $\mathbf{5}_\mathrm{H}$ and $\mathbf{45}_\mathrm{H}$ acquire their VEVs, the mass terms for fermions become
\begin{eqnarray}
  \mathcal{L} & \supset & - M_{u, i j} \overline{u_{i \mathrm{L}}} u_{j \mathrm{R}}
  - M_{d, i j} \overline{d_{i \mathrm{L}}} d_{j \mathrm{R}} - M_{\ell, i j} 
  \overline{\ell_{i \mathrm{L}}} \ell_{j \mathrm{R}}\\
  &  & - \frac{1}{2} \begin{pmatrix}
    \overline{\nu_{\mathrm{L}}} & \overline{(N_{\mathrm{R}})^{\mathrm{C}}}
  \end{pmatrix} \begin{pmatrix}
    \mathbf{0} & M_\mathrm{D}\\
    M_\mathrm{D}^\mathrm{T} & M_N
  \end{pmatrix} \begin{pmatrix}
    (\nu_{\mathrm{L}})^{\mathrm{C}}\\
    N_{\mathrm{R}}
  \end{pmatrix} + \mathrm{H.c.},
\end{eqnarray}
with
\begin{eqnarray}
  M_{u, i j} &\equiv& \sqrt{2} v_5 (Y_{j i}^{10 \ast} + Y_{i j}^{10 \ast}) + 2
  \sqrt{2} v_{45} (\tilde{Y}_{j i}^{45} - \tilde{Y}_{i j}^{45}),
\\
  M_{d, i j} &\equiv& \frac{1}{2} Y_{i j}^{5 \ast} v_5 + Y_{i j}^{45 \ast}
  v_{45},
  \\
  M_{\ell, i j} &\equiv& \frac{1}{2} Y_{j i}^{5 \ast} v_5 - 3 Y_{j
  i}^{45 \ast} v_{45},
\\
M_{\mathrm{D}} &\equiv& \frac{v_5}{\sqrt{2}} Y^{1 \ast}.
\end{eqnarray}

For simplicity, we assume that the Yukawa coupling matrices $Y^5$, $Y^{45}$, and $\tilde{Y}^{45}$ are symmetric.
Consequently, the mass matrices $M_u$, $M_d$, and $M_\ell$ are also symmetric and can be diagonalized by unitary matrices $U_u$, $U_d$, and $U_\ell$, respectively.
They are related by
\begin{equation}
  M_u = U_u \tilde{M}_u U^{\mathrm{T}}_u, \quad M_d = U_d \tilde{M}_d
  U^{\mathrm{T}}_d, \quad M_{\ell} = U_{\ell} \tilde{M}_{\ell}
  U^{\mathrm{T}}_{\ell},
\end{equation}
where $\tilde{M}_u = \operatorname{diag} (m_u, m_c, m_t)$, $\tilde{M}_d = \operatorname{diag} (m_d, m_s, m_b)$, and $\tilde{M}_{\ell} = \operatorname{diag} (m_e, m_{\mu},  m_{\tau})$ are the diagonalized mass matrices.
Hence, the mass eigenstates for quarks and charged leptons are given by $u^\mathrm{m}_{i \mathrm{L}} \equiv (U^{\dag}_u)_{i j} u_{j \mathrm{L}}$, $u^\mathrm{m}_{i
\mathrm{R}} \equiv (U^{\mathrm{T}}_u)_{i j} u_{j \mathrm{R}}$, $d^\mathrm{m}_{i \mathrm{L}} \equiv (U^{\dag}_d)_{i j} d_{j \mathrm{L}}$, $d^\mathrm{m}_{i
\mathrm{R}} \equiv (U^{\mathrm{T}}_d)_{i j} d_{j \mathrm{R}}$, $\ell^\mathrm{m}_{i \mathrm{L}} \equiv (U^{\dag}_{\ell})_{i j} \ell_{j \mathrm{L}}$, and $\ell^\mathrm{m}_{i \mathrm{R}} \equiv (U^{\mathrm{T}}_{\ell})_{i j} \ell_{j
\mathrm{R}}$.
The Cabibbo-Kobayashi-Maskawa (CKM) matrix \cite{Cabibbo:1963yz,Kobayashi:1973fv} is defined as $V_{\mathrm{CKM}} \equiv U_u^{\dag} U_d$.
Conversely, the Yukawa coupling matrices are expressed as
\begin{eqnarray}
  Y^{10 \dag} + Y^{10 \ast} &=& \frac{1}{\sqrt{2} v_5} U_u \tilde{M}_u
  U^{\mathrm{T}}_u,
\label{eq:Y10}
\\
  Y^{5 \ast} &=& \frac{1}{2 v_5} (3 U_d \tilde{M}_d U^{\mathrm{T}}_d + U_{\ell}
  \tilde{M}_{\ell} U^{\mathrm{T}}_{\ell}),
\label{eq:Y5}
\\
Y^{45 \ast} &=& \frac{1}{4
  v_{45}} (U_d \tilde{M}_d U^{\mathrm{T}}_d - U_{\ell} \tilde{M}_{\ell}
  U^{\mathrm{T}}_{\ell}).
\end{eqnarray}

In addition, we assume the seesaw hierarchy $M_N \gg M_{\mathrm{D}}$, so that the mass matrix for active neutrinos, $M_\nu$, satisfies the seesaw relation
\begin{equation}
M_\nu = - M_{\mathrm{D}} M_N^{- 1} M_{\mathrm{D}}^{\mathrm{T}}.
\end{equation}
Both $M_\nu$ and $M_N$ can be diagonalized by unitary matrices $U_\nu$ and $U_N$, respectively, such that
\begin{equation}
  U_{\nu} \tilde{M}_{\nu} U_{\nu}^{\mathrm{T}} = M_{\nu}, \quad U_N
  \tilde{M}_N U_N^{\mathrm{T}} = M_N,
\end{equation}
where $\tilde{M}_{\nu} = \operatorname{diag} (m_{\nu_1}, m_{\nu_2}, m_{\nu_3})$ and $\tilde{M}_N =
\operatorname{diag} (m_{N_1}, m_{N_2}, m_{N_3})$ are the diagonalized mass matrices for light and heavy Majorana neutrinos, respectively.
The mass eigenstates of neutrinos are defined by  $(\nu^{\mathrm{m}}_{\mathrm{L}})^{\mathrm{C}} \equiv U_{\nu}^{\mathrm{T}}
(\nu_{\mathrm{L}})^{\mathrm{C}}$ and $N^{\mathrm{m}}_{\mathrm{R}} \equiv
U_N^{\mathrm{T}} N_{\mathrm{R}}$.
The corresponding Majorana neutrino fields are then constructed as $\nu^{\mathrm{m}}_i \equiv \nu^{\mathrm{m}}_{i \mathrm{L}} +
(\nu^{\mathrm{m}}_{i \mathrm{L}})^{\mathrm{C}}$ and $N^{\mathrm{m}}_i \equiv
N^{\mathrm{m}}_{i \mathrm{R}} + (N^{\mathrm{m}}_{i \mathrm{R}})^{\mathrm{C}}$.
The Pontecorvo-Maki-Nakagawa-Sakata (PMNS) matrix \cite{Maki:1962mu,Pontecorvo:1967fh} is defined as $V_{\mathrm{PMNS}} \equiv U_{\ell}^{\dag} U_{\nu}$.

From the Lagrangian \eqref{eq:L:Yukawa_Mass}, we extract the Yukawa couplings of the scalar leptoquarks $\Delta_a$ with SM fermions as
\begin{eqnarray}
\mathcal{L} &\supset&
- 2 (Y_{i j}^{10} + Y_{j i}^{10}) \overline{\ell_{i \mathrm{R}}} (u_{a j
\mathrm{R}})^{\mathrm{C}} \Delta_a + 2 (Y_{i j}^{10} + Y_{j i}^{10})
\varepsilon_{a b c} \overline{(d_{a i \mathrm{L}})^{\mathrm{C}}} u_{b j
\mathrm{L}} \Delta_c
\nonumber\\
&& + \frac{Y_{i j}^5}{\sqrt{2}}\, \varepsilon_{a b c} \overline{u_{a i \mathrm{R}}}
(d_{b j \mathrm{R}})^{\mathrm{C}} \Delta_c^{\ast} + \frac{Y_{i j}^5}{\sqrt{2}}\,
\overline{(u_{a i \mathrm{L}})^{\mathrm{C}}} \ell_{j \mathrm{L}}
\Delta_a^{\ast} - \frac{Y_{i j}^5}{\sqrt{2}} \,\overline{(d_{a i
\mathrm{L}})^{\mathrm{C}}} \nu_{j \mathrm{L}} \Delta_a^{\ast}
+ \mathrm{H.c.},
\label{eq:L_Delta_Yuk}
\end{eqnarray}
where $a,b,c=1,2,3$ are color indices.

\section{Proton Decay}
\label{sub:proton_decay}

The scalar leptoquarks $\Delta_a$ can mediate proton decay.
This process is conveniently discussed in the framework of effective field theory, with effective operators simultaneously violating baryon number $B$ and lepton number $L$ by one unit.
To the lowest order, the relevant dimension-6 effective operators involving three quarks and one lepton, induced by the exchange of $\Delta_a$, can be written as \cite{Nath:2006ut}
\begin{eqnarray}
  \mathcal{L}_{\mathrm{eff}} & \supset &  a (d_{i \mathrm{L}},
  \ell_{j \mathrm{L}}) \varepsilon_{a b c} \overline{(u^{\mathrm{m}}_{a 1
  \mathrm{L}})^{\mathrm{C}}} d^{\mathrm{m}}_{b i \mathrm{L}} 
  \overline{(u^{\mathrm{m}}_{c 1 \mathrm{L}})^{\mathrm{C}}}
  \ell^{\mathrm{m}}_{j \mathrm{L}}
   - a (d_{i \mathrm{L}}, \ell_{j \mathrm{R}}) \varepsilon_{a b c}
  \overline{(u^{\mathrm{m}}_{a 1 \mathrm{L}})^{\mathrm{C}}} d^{\mathrm{m}}_{b
  i \mathrm{L}} \overline{(u^{\mathrm{m}}_{c 1 \mathrm{R}})^{\mathrm{C}}}
  \ell_{j \mathrm{R}}^{\mathrm{m}} \nonumber\\
  &  & - a (d_{i \mathrm{R}}, \ell_{j \mathrm{L}}) \varepsilon_{a b c}
  \overline{(d^{\mathrm{m}}_{a i \mathrm{R}})^{\mathrm{C}}} u^{\mathrm{m}}_{b
  1 \mathrm{R}} \overline{(u^{\mathrm{m}}_{c 1 \mathrm{L}})^{\mathrm{C}}}
  \ell^{\mathrm{m}}_{j \mathrm{L}}
   + a (d_{i \mathrm{R}}, \ell_{j \mathrm{R}}) \varepsilon_{a b c}
  \overline{(d^{\mathrm{m}}_{a i \mathrm{R}})^{\mathrm{C}}} u^{\mathrm{m}}_{b
  1 \mathrm{R}} \overline{(\ell^{\mathrm{m}}_{j \mathrm{R}})^{\mathrm{C}}}
  u^{\mathrm{m}}_{c 1 \mathrm{R}} \nonumber\\
  &  & + a (d_{i \mathrm{L}}, d_{j \mathrm{L}}, \nu_{k \mathrm{L}})
  \varepsilon_{a b c} \overline{(u^{\mathrm{m}}_{a 1
  \mathrm{L}})^{\mathrm{C}}} d^{\mathrm{m}}_{b i \mathrm{L}}
  \overline{(d^{\mathrm{m}}_{c j \mathrm{L}})^{\mathrm{C}}}
  \nu^{\mathrm{m}}_{k \mathrm{L}}
  - a (d_{i \mathrm{R}}, d_{j \mathrm{L}}, \nu_{k \mathrm{L}})
  \varepsilon_{a b c} \overline{(d^{\mathrm{m}}_{a i
  \mathrm{R}})^{\mathrm{C}}} u^{\mathrm{m}}_{b 1 \mathrm{R}}
  \overline{(d^{\mathrm{m}}_{c j \mathrm{L}})^{\mathrm{C}}}
  \nu^{\mathrm{m}}_{k \mathrm{L}}.
  \nonumber\\*
\end{eqnarray}
Based on the Yukawa couplings in the Lagrangian~\eqref{eq:L_Delta_Yuk}, the effective coefficients above are obtained as \cite{,Dorsner:2012nq}
\begin{eqnarray}
  a (d_{i \mathrm{L}}, \ell_{j \mathrm{L}}) & = & -
  \frac{\sqrt{2}}{m_{\Delta}^2} [U_u^{\mathrm{T}} (Y^{10} + Y^{10 \mathrm{T}})
  U_d]_{1 i} (U_u^{\mathrm{T}} Y^5 U_{\ell})_{1 j} 
\label{eq:a_diL_ljL}
\nonumber\\
  & = & - \frac{1}{2 m_{\Delta}^2 v_5^2} (\tilde{M}_u V_{\mathrm{CKM}})_{1 i}
  (3 V_{\mathrm{CKM}}^{\ast} \tilde{M}_d V_{\mathrm{CKM}}^{\dag} U_{u
  \ell}^{\ast} + U_{u \ell} \tilde{M}_{\ell})_{1 j}, \\
  a (d_{i \mathrm{L}}, \ell_{j \mathrm{R}}) & = & - \frac{4}{m_{\Delta}^2}
  [U_u^{\mathrm{T}} (Y^{10} + Y^{10 \mathrm{T}}) U_d]_{1 i} [U^{\dag}_{\ell}
  (Y^{10 \dag} + Y^{10 \ast}) U^{\ast}_u]_{j 1} \nonumber\\
  & = & - \frac{2}{m_{\Delta}^2 v^2_5} (\tilde{M}_u V_{\mathrm{CKM}})_{1 i}
  (U_{u \ell}^{\mathrm{T}} \tilde{M}_u)_{j 1}, \\
  a (d_{i \mathrm{R}}, \ell_{j \mathrm{L}}) & = & \frac{1}{2 m_{\Delta}^2}
  (U^{\dag}_d Y^{5 \dag} U_u^{\ast})_{i 1} (U_u^{\mathrm{T}} Y^5 U_{\ell})_{1
  j} \nonumber\\
  & = & \frac{1}{8 m_{\Delta}^2 v_5^2} (3 \tilde{M}_d
  V_{\mathrm{CKM}}^{\mathrm{T}} + V_{\mathrm{CKM}}^{\dag} U_{u \ell}^{\ast}
  \tilde{M}_{\ell} U_{u \ell}^{\dag})_{i 1} \text{} (3 V_{\mathrm{CKM}}^{\ast}
  \tilde{M}_d V_{\mathrm{CKM}}^{\dag} U_{u \ell}^{\ast} + U_{u \ell}
  \tilde{M}_{\ell})_{1 j},\qquad \\
  a (d_{i \mathrm{R}}, \ell_{j \mathrm{R}}) & = &
  \frac{\sqrt{2}}{m_{\Delta}^2} (U^{\dag}_d Y^{5 \dag} U_u^{\ast})_{i 1}
  [U^{\dag}_{\ell} (Y^{10 \dag} + Y^{10 \ast}) U^{\ast}_u]_{j 1} \nonumber\\
  & = & \frac{1}{2 m_{\Delta}^2 v_5^2} (3 \tilde{M}_d
  V_{\mathrm{CKM}}^{\mathrm{T}} + V_{\mathrm{CKM}}^{\dag} U_{u \ell}^{\ast}
  \tilde{M}_{\ell} U_{u \ell}^{\dag})_{i 1} (U_{u \ell}^{\mathrm{T}}
  \tilde{M}_u)_{j 1}, \\
  a (d_{i \mathrm{L}}, d_{j \mathrm{L}}, \nu_{k \mathrm{L}}) & = &
  \frac{\sqrt{2}}{m_{\Delta}^2} [U_u^{\mathrm{T}} (Y^{10} + Y^{10 \mathrm{T}})
  U_d]_{1 i} (U_d^{\mathrm{T}} Y^5 U_{\nu})_{j k} \nonumber\\
  & = & \frac{1}{2 m_{\Delta}^2 v_5^2} (\tilde{M}_u V_{\mathrm{CKM}})_{1 i}
  ( 3 \tilde{M}_d V_{\mathrm{CKM}}^{\dag} U_{u \ell}^{\ast}
  V_{\mathrm{PMNS}} + V_{\mathrm{CKM}}^{\mathrm{T}} U_{u \ell}
  \tilde{M}_{\ell} V_{\mathrm{PMNS}} )_{j k}, \\
  a (d_{i \mathrm{R}}, d_{j \mathrm{L}}, \nu_{k \mathrm{L}}) & = & -
  \frac{1}{2 m_{\Delta}^2} (U^{\dag}_d Y^{5 \dag} U_u^{\ast})_{i 1}
  (U_d^{\mathrm{T}} Y^5 U_{\nu})_{j k} \nonumber\\
  & = & - \frac{1}{8 m_{\Delta}^2 v_5^2} (3 \tilde{M}_d
  V_{\mathrm{CKM}}^{\mathrm{T}} + V_{\mathrm{CKM}}^{\dag} U_{u \ell}^{\ast}
  \tilde{M}_{\ell} U_{u \ell}^{\dag})_{i 1} \nonumber\\
  &  & \quad \times ( 3 \tilde{M}_d V_{\mathrm{CKM}}^{\dag} U_{u
  \ell}^{\ast} V_{\mathrm{PMNS}} + V_{\mathrm{CKM}}^{\mathrm{T}} U_{u \ell}
  \tilde{M}_{\ell} V_{\mathrm{PMNS}} )_{j k},
\label{eq:a_diR_djL_nukL}
\end{eqnarray}
where $m_\Delta$ is the common mass of the three scalar leptoquarks $\Delta_a$, and $U_{u \ell} \equiv U_u^{\mathrm{T}} U_{\ell}^{\ast}$ is a combination of two mixing matrices.

Using the chiral Lagrangian techniques, partial widths for proton decay channels $p \rightarrow \ell_i^+ \pi^0$ ($i=1,2$), $p \rightarrow \pi^+ \bar{\nu}_i$ ($i=1,2,3$), and $p \rightarrow K^+ \bar{\nu}_i$ ($i=1,2,3$) are given by \cite{Nath:2006ut}
\begin{eqnarray}
  \Gamma_{p \rightarrow \ell_i^+ \pi^0} & = & \frac{(m_p^2 -
  m_{\pi^0}^2)^2}{64 \pi f_{\pi}^2 m_p^3} (1 + D + F)^2 \nonumber\\
  &  & \times \big[| \alpha a (d_{1 \mathrm{L}}, \ell_{i \mathrm{L}}) + \beta a
  (d_{1 \mathrm{R}}, \ell_{i \mathrm{L}}) |^2 + | \alpha a (d_{1 \mathrm{L}},
  \ell_{i \mathrm{R}}) + \beta a (d_{1 \mathrm{R}}, \ell_{i \mathrm{R}}) |^2\big],
\\
\Gamma_{p \rightarrow \pi^+ \bar{\nu}_i} & = & 
\frac{(m_p^2 - m^2_{\pi^+})^2}{32 \pi f_{\pi}^2 m_p^3} (1 + D + F)^2 \left|
\alpha a (d_{1 \mathrm{R}}, d_{1 \mathrm{L}}, \nu_{i \mathrm{L}}) + \beta a
(d_{1 \mathrm{L}}, d_{1 \mathrm{L}}, \nu_{i \mathrm{L}}) \right|^2,
\\
\Gamma_{p \rightarrow K^+ \bar{\nu}_i} & = & 
\frac{(m_p^2 - m^2_{K^+})^2}{32 \pi f_{\pi}^2 m_p^3} \Big| \beta a (d_{1
\mathrm{L}}, d_{2 \mathrm{L}}, \nu_{i \mathrm{L}}) + \alpha a (d_{1
\mathrm{R}}, d_{2 \mathrm{L}}, \nu_{i \mathrm{L}}) 
\nonumber\\
&& \quad - \frac{{m_p} }{2 m_{\Sigma^0}} [\beta a (d_{2 \mathrm{L}}, d_{1 \mathrm{L}},
\nu_{i \mathrm{L}}) + \alpha a (d_{2 \mathrm{R}}, d_{1 \mathrm{L}}, \nu_{i
\mathrm{L}})] (D - F)
\nonumber\\
&& \quad + \frac{m_p}{6 m_{\Lambda}} \big\{ \beta a (d_{2 \mathrm{L}}, d_{1 \mathrm{L}},
\nu_{i \mathrm{L}}) + \alpha a (d_{2 \mathrm{R}}, d_{1 \mathrm{L}}, \nu_{i
\mathrm{L}}) 
\nonumber\\
&& \qquad\qquad + 2 [\beta a (d_{1 \mathrm{L}}, d_{2 \mathrm{L}}, \nu_{i
\mathrm{L}}) + \alpha a (d_{1 \mathrm{R}}, d_{2 \mathrm{L}}, \nu_{i
\mathrm{L}})] \big\} (D + 3 F) \Big|^2.
\end{eqnarray}
Here, $m_p = 938.27208816(29)~\si{MeV}$, $m_{\pi^0} = 134.9768(5)~\si{MeV}$, $m_{\pi^+} = 139.57039(18)~\si{MeV}$, $m_{K^+} = 493.677(15)~\si{MeV}$, $m_\Lambda = 1115.683(6)~\si{MeV}$, and $m_{\Sigma^0} = 1192.642(24)~\si{MeV}$ denote the masses of the proton, $\pi^0$ meson, $\pi^+$ meson, $K^+$ meson, $\Lambda$ baryon, $\Sigma^0$ baryon, respectively \cite{ParticleDataGroup:2024cfk}.
$f_\pi = 130.2(1.2)~\si{MeV}$ is the pion decay constant \cite{ParticleDataGroup:2024cfk}.
$\alpha = -0.0112(25)~\si{GeV^3}$ and $\beta = 0.0120(26)~\si{GeV^3}$ are the coefficients in the matrix elements of the three quark operators between the proton and vacuum states, while measurements of the nucleon axial charge and form factors for semileptonic hyperon decay give $D = 0.80(1)$ and $F = 0.47(1)$ \cite{Aoki:2008ku}.

For the above proton decay channels, the experimental lower limits on the partial lifetimes at $90\%$ confidence level (CL) are $\tau_{p \to e^+ \pi^0} > \num{2.4e34}~\si{yr}$, $\tau_{p \to \mu^+ \pi^0} > \num{1.6e34}~\si{yr}$, $\tau_{p \to \pi^+ \bar{\nu}} > \num{3.9e32}~\si{yr}$, and $\tau_{p \to K^+ \bar{\nu}} > \num{5.9e33}~\si{yr}$ \cite{ParticleDataGroup:2024cfk}.
Converting these limits into partial decay widths yields the following upper bounds:
$\Gamma_{p \to e^+ \pi^0} < \num{8.7e-67}~\si{GeV}$,
$\Gamma_{p \to \mu^+ \pi^0} < \num{1.3e-66}~\si{GeV}$,
$\Gamma_{p \to \pi^+ \bar{\nu}} < \num{5.3e-65}~\si{GeV}$,
and $\Gamma_{p \to K^+ \bar{\nu}} < \num{3.5e-66}~\si{GeV}$.

Below, we perform a random scan over the parameter space to study the constraints from proton decay.
Before that, we need to parametrize the relevant mixing matrices.
Any $3\times 3$ unitary matrix $U$ can be parametrized as \cite{Rasin:1997pn}
\begin{equation}
  U = K (\rho_1, \rho_2) R_{23} (\theta_{23}) R_{13} (\theta_{13}, \delta)
  R_{12} (\theta_{12}) D (\eta_1, \eta_2, \eta_3),
\end{equation}
where $K (\rho_1, \rho_2) = \operatorname{diag} (1, \mathrm{e}^{\mathrm{i} \rho_1}, \mathrm{e}^{\mathrm{i}
\rho_2})$ and $D(\eta_1, \eta_2, \eta_3) = \operatorname{diag} (\mathrm{e}^{\mathrm{i} \eta_1}, \mathrm{e}^{\mathrm{i} \eta_2},
\mathrm{e}^{\mathrm{i} \eta_3})$ are two diagonal matrices together containing five independent phases, and 
\begin{eqnarray}
  R_{23} (\theta_{23}) & = & \begin{pmatrix}
    1 &  & \\
    & c_{23} & s_{23}\\
    & - s_{23} & c_{23}
  \end{pmatrix},
\\
  R_{13} (\theta_{13}, \delta) & = & \begin{pmatrix}
    c_{13} &  & s_{13} \mathrm{e}^{- \mathrm{i} \delta}\\
    & 1 & \\
    - s_{13} \mathrm{e}^{\mathrm{i} \delta} &  & c_{13}
  \end{pmatrix},
\\
  R_{12} (\theta_{12}) & = & \begin{pmatrix}
    c_{12} & s_{12} & \\
    - s_{12} & c_{12} & \\
    &  & 1
  \end{pmatrix} 
\end{eqnarray}
involve three mixing angles and one phase.
Here, we use the shorthand notations $c_{i j} \equiv \cos \theta_{i j}$ and $s_{i j} \equiv \sin \theta_{i j}$.

Since five independent phases in the CKM matrix can be absorbed by redefining the quark fields, it reduces to the form of 
\begin{equation}
V_{\mathrm{CKM}} = R_{23} (\theta^{\mathrm{CKM}}_{23}) R_{13}(\theta^{\mathrm{CKM}}_{13}, \delta^{\mathrm{CKM}}) R_{12} (\theta^{\mathrm{CKM}}_{12}).
\end{equation}
The global fit of flavor data gives $\sin\theta^\mathrm{CKM}_{12} = 0.22501(68)$, $\sin\theta^\mathrm{CKM}_{23} = 0.04183^{+0.00079}_{-0.00069}$, $\sin\theta^\mathrm{CKM}_{13} = 0.003732^{+0.000090}_{-0.000085}$, and $\delta^\mathrm{CKM} = 1.147(26)$ \cite{ParticleDataGroup:2024cfk}.
For the PMNS matrix, three independent phases can be removed by redefinitions of the charged lepton fields, and it takes the form
\begin{equation}
V_{\mathrm{PMNS}} = R_{23} (\theta^{\mathrm{PMNS}}_{23}) R_{13} (\theta^{\mathrm{PMNS}}_{13}, \delta^{\mathrm{PMNS}}) R_{12}
(\theta^{\mathrm{PMNS}}_{12}) D (\eta^{\mathrm{PMNS}}_1, \eta^{\mathrm{PMNS}}_2, 0).
\end{equation}
The global fit of neutrino oscillation data assuming normal mass ordering $ m_{\nu_1} < m_{\nu_2} <  m_{\nu_3}$ leads to $\theta^\mathrm{PMNS}_{12}/^\circ = 33.41^{+0.75}_{-0.72}$, $\theta^\mathrm{PMNS}_{23}/^\circ = 49.1^{+1.0}_{-1.3}$, $\theta^\mathrm{PMNS}_{13}/^\circ = 8.54^{+0.11}_{-0.12}$, $\delta^\mathrm{PMNS}/^\circ = 197^{+42}_{-25}$, $\Delta m^2_{21} \equiv m_{\nu_2}^2 - m_{\nu_1}^2 = 7.41^{+0.21}_{-0.20}\times 10^{-5}~\si{eV^2}$ and $\Delta m^2_{32} \equiv m_{\nu_3}^2 - m_{\nu_2}^2 = 2.437^{+0.028}_{-0.027}\times 10^{-3}~\si{eV^2}$  \cite{ParticleDataGroup:2024cfk}.
In the remaining mixing matrices, we take $U_u$ and $U_\ell$ as two independent matrices and parametrize them as
\begin{eqnarray}
  U_u & = & K (\rho^u_1, \rho^u_2) R_{23} (\theta^u_{23}) R_{13}
  (\theta^u_{13}, \delta^u) R_{12} (\theta^u_{12}) D (\eta^u_1, \eta^u_2,
  \eta^u_3),
\\
  U_{\ell} & = & K (\rho^{\ell}_1, \rho^{\ell}_2) R_{23} (\theta^{\ell}_{23})
  R_{13} (\theta^{\ell}_{13}, \delta^{\ell}) R_{12} (\theta^{\ell}_{12}) D
  (\eta^{\ell}_1, \eta^{\ell}_2, \eta^{\ell}_3).
\end{eqnarray}
Furthermore, the $\mathbf{5}_\mathrm{H}$ VEV $v_5$ and the scalar leptoquark mass $m_\Delta$ are treated as independent parameters.

The partial proton decay widths also depend on the masses of the elementary fermions.
Their measured values are $m_e = 0.51099895000(15)~\si{MeV}$, $m_\mu = 105.6583755(23)~\si{MeV}$, $m_\tau = 1776.93(9)~\si{MeV}$, $m_u = 2.16(7)~\si{MeV}$, $m_d = 4.70(7)~\si{MeV}$, $m_s = 93.5(8)~\si{MeV}$, $m_c = 1.2730(46)~\si{GeV}$, $m_b = 4.183(7)~\si{GeV}$, and $m_t = 172.57(29)~\si{GeV}$ \cite{ParticleDataGroup:2024cfk}.
For the active neutrino masses, we assume normal ordering and set $m_{\nu_1} = 0$ for simplicity.

\begin{figure}[!t]
\centering
\subfigure[$p \rightarrow e^+ \pi^0$\label{fig:gamma_p_decay_epi}]{\includegraphics[width=.48\textwidth]{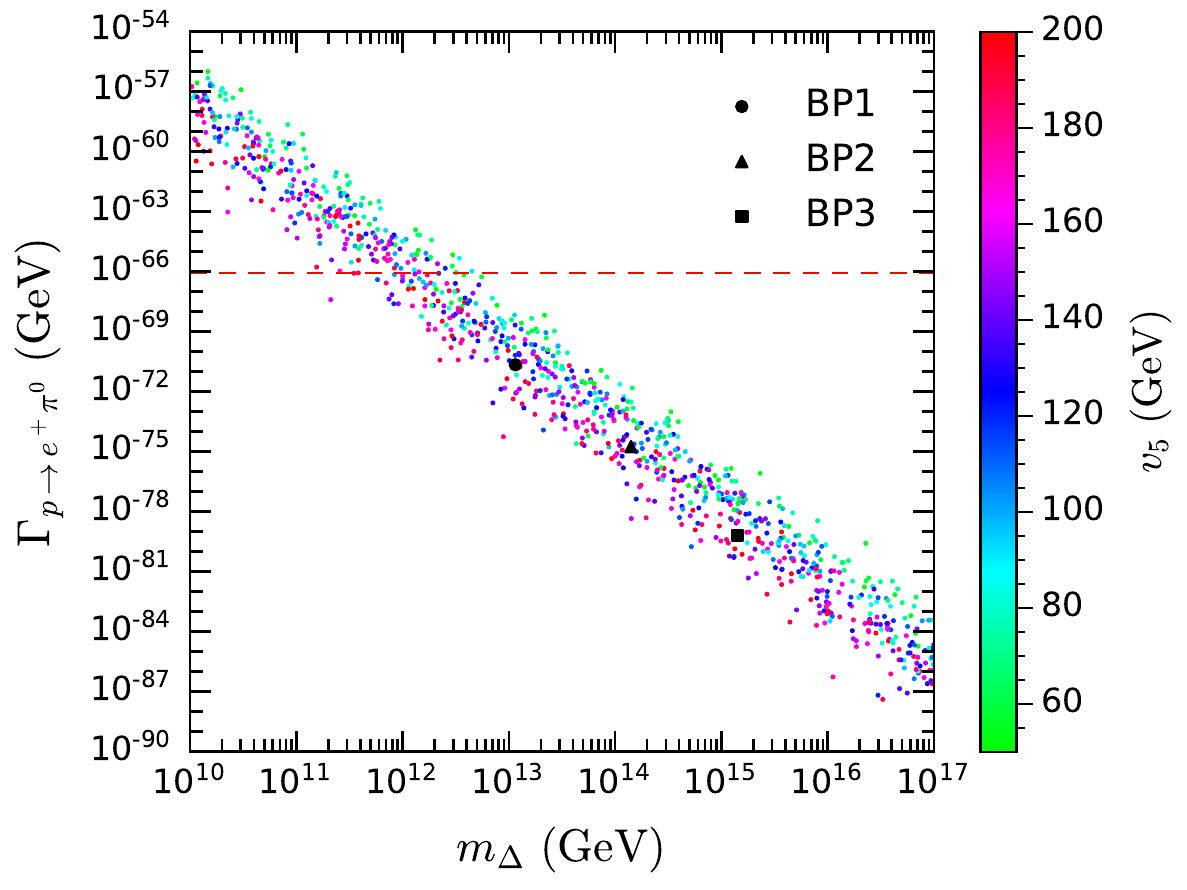}}
\hspace{.02\textwidth}
\subfigure[$p\rightarrow \mu^+ \pi^0$\label{fig:gamma_p_decay_mupi}]{\includegraphics[width=.48\textwidth]{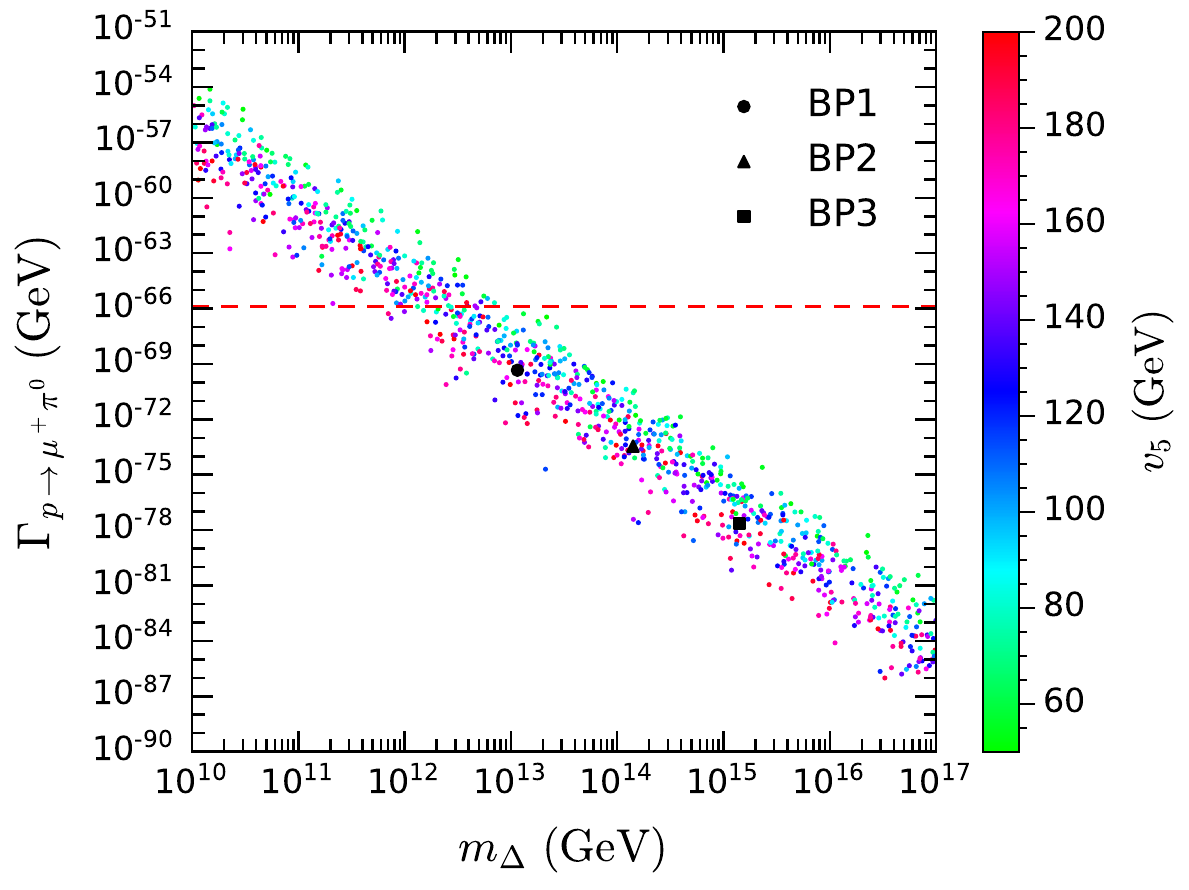}}
\subfigure[$p\rightarrow K^+ \bar{\nu}$\label{fig:gamma_p_decay_e}]{\includegraphics[width=.48\textwidth]{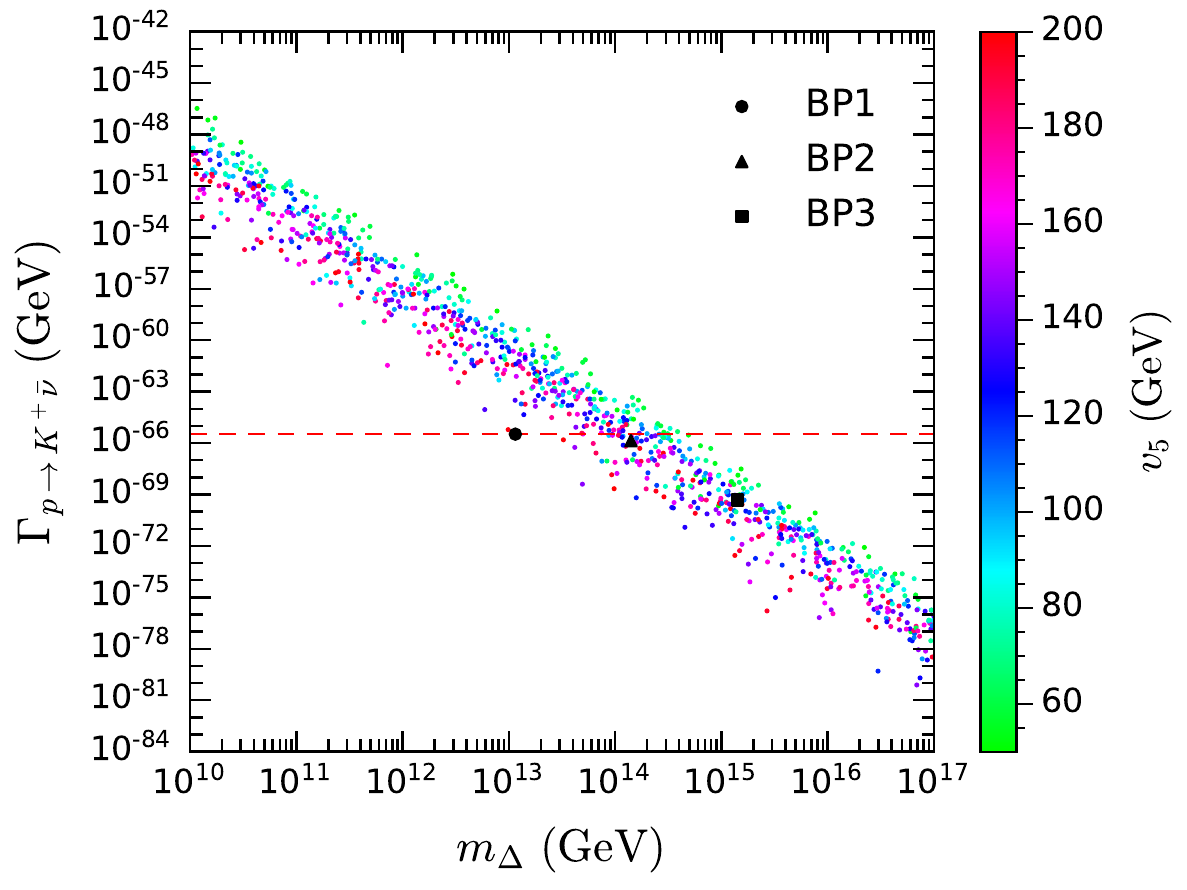}}
\hspace{.02\textwidth}
\subfigure[$p\rightarrow \pi^+ \bar{\nu}$\label{fig:gamma_p_decay_mu}]{\includegraphics[width=.48\textwidth]{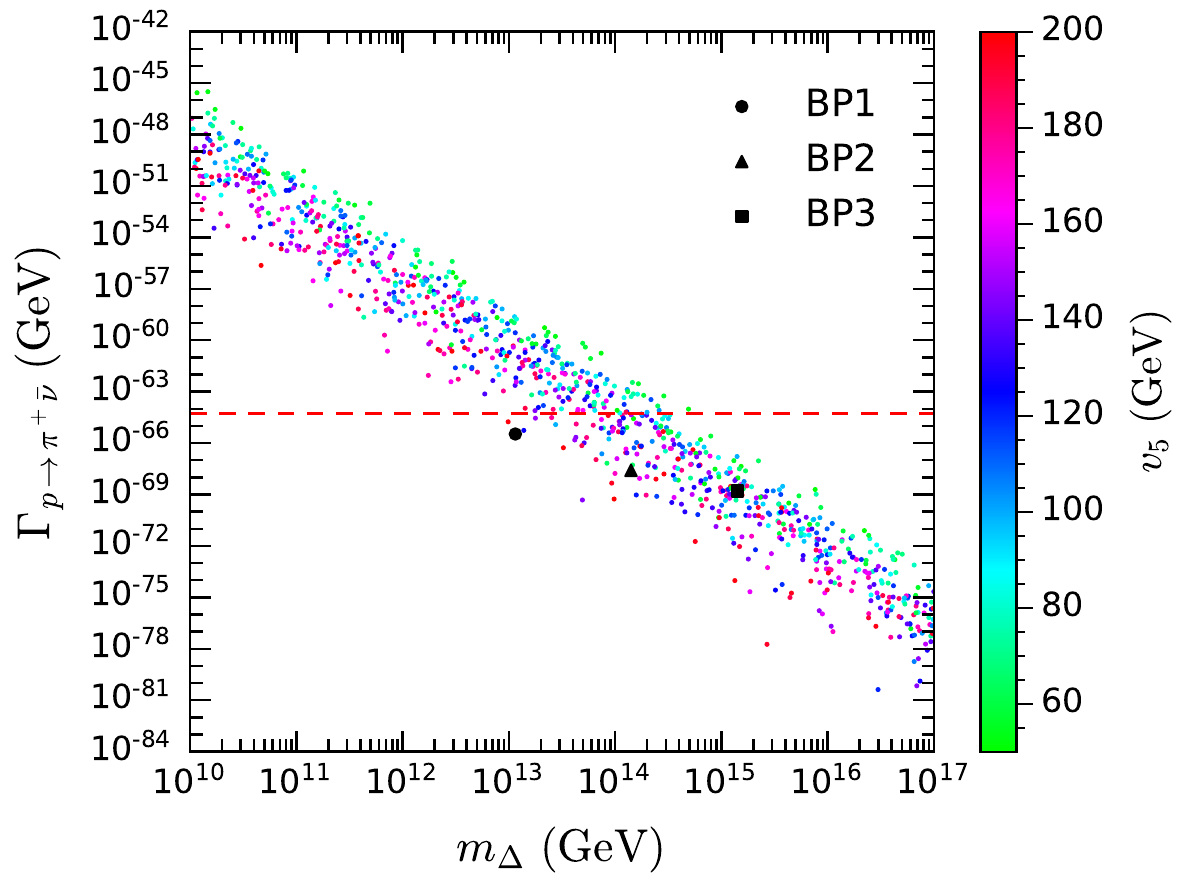}}
\caption{Partial widths for proton decay channels $p\rightarrow e^+ \pi^0$ (a), $p\rightarrow \mu^+ \pi^0$ (b), $p\rightarrow K^+ \bar{\nu}$ (c), and $p\rightarrow \pi^+ \bar{\nu}$ (d).
The red dashed lines indicates the experimental upper limits at 90\% CL.
The black dot, triangle, and square mark three benchmark points.}
\label{fig:proton_decay}
\end{figure}

We fix the measured parameters to their mean or best-fit values, and carry out a random scan over the free parameters within the following ranges:
\begin{alignat}{2}
& 0 < \theta^u_{ij}, \theta^\ell_{ij} < \frac{\pi}{2},&\qquad
& 0 < \eta^\mathrm{PMNS}_i, \eta^u_i, \rho^u_i, \delta^u, \eta^\ell_i, \rho^\ell_i, \delta^\ell < 2\pi,
\nonumber\\
& 50~\si{GeV} < v_5 < 200~\si{GeV},&\qquad
& 10^{10}~\si{GeV} < m_\Delta < 10^{17}~\si{GeV}.
\end{alignat}
For each parameter points, we calculate the predicted partial widths for the proton decay channels $p\rightarrow e^+ \pi^0$, $p\rightarrow \mu^+ \pi^0$, $p\rightarrow K^+ \bar{\nu}$, and $p\rightarrow \pi^+ \bar{\nu}$, where the latter two channels involve an implicit summation over the three active neutrino families.
These results are shown in Fig.~\ref{fig:proton_decay}, where the red dashed lines denote the experimental upper limits and the color axes indicate the values of $v_5$.
Because the effective coefficients \eqref{eq:a_diL_ljL}--\eqref{eq:a_diR_djL_nukL} scale as $(m_\Delta v_5)^{-2}$, the predicted partial widths are negatively correlated with both $m_\Delta$ and $v_5$, a trend clearly reflected by the distribution of the parameter points.
The strongest constraint arises from the $p\rightarrow K^+ \bar{\nu}$ channel, which excludes parameter points with $m_\Delta \lesssim 10^{13}~\si{GeV}$.

Among the parameter points that satisfy the experimental bounds on proton decay, we select three benchmark points (BPs), denoted as BP1, BP2, and BP3, for further study.
They are indicated in Fig.~\ref{fig:proton_decay} and their parameter values are listed in Table~\ref{tab:BPs}.
The scalar leptoquark mass $m_\Delta$ in BP1, BP2, and BP3 are $\num{1.15e13}$, $\num{1.41e14}$, and $\num{1.42e15}~\si{GeV}$, respectively.

\begin{table}[!t]
\centering
\setlength\tabcolsep{1em}
\renewcommand{\arraystretch}{1.5}
\caption{Parameters in three benchmark points.}
\label{tab:BPs}
\begin{tabular}{cccc}
\hline\hline
 & BP1 & BP2 & BP3 \\
\hline
$m_\Delta~ (\mathrm{GeV})$ & $1.15\times 10^{13} $ & $1.41\times 10^{14}$ & $1.42\times 10^{15}$ \\
$v_5~(\mathrm{GeV})$ & $164.75$ & $106.43$ & $148.58$\\
$\eta^{\mathrm{PMNS}}_1,\eta^{\mathrm{PMNS}}_2$ & $4.76,~ 5.13$ & $1.23,~ 4.47$ & $6.20,~ 3.13$ \\
$\theta_{12}^u, \theta_{13}^u, \theta_{23}^u$& $1.52,~ 0.25,~ 1.53$ & $1.32,~ 0.21,~ 0.91$ & $1.47,~ 0.64,~ 0.42$\\
$\delta^u, \rho_{1}^u, \rho_{2}^u$& $1.54,~ 2.75,~ 4.43$ & $0.12,~ 0.33,~ 1.53$ & $2.66,~ 1.50,~ 2.73$ \\
$\eta_{1}^u, \eta_{2}^u, \eta_{3}^u$& $1.67,~ 3.21,~ 1.05$ & $5.58,~ 0.26,~ 4.34$ & $3.62,~ 5.91,~ 2.04$ \\
$\theta_{12}^\ell, \theta_{13}^\ell, \theta_{23}^\ell$ & $1.17,~ 0.25,~ 0.71$ & $1.35,~ 1.35,~ 0.18$ & $0.34,~ 0.10,~ 0.03$\\
$\delta^\ell, \rho_{1}^\ell, \rho_{2}^\ell$ & $4.26,~ 3.66,~ 1.80$ & $1.85,~ 0.26,~ 2.45$ & $5.01,~ 1.47,~ 4.82$\\
$\eta_{1}^\ell, \eta_{2}^\ell, \eta_{3}^\ell$ & $5.21,~ 5.67,~ 2.35$ & $6.10,~ 0.78,~ 1.07$ & $1.97,~ 1.48,~ 6.00$ \\
\hline\hline
\end{tabular}
\end{table}

\section{Number density of scalar leptoquarks}
\label{sub:num_den}

In order to estimate the stochastic GWs arising from graviton bremsstrahlung in decays of the scalar leptoquarks $\Delta_a$ and the antiparticles $\bar{\Delta}_a$, we need to track the evolution of their number densities $n_{\Delta_a}$ and $n_{\bar{\Delta}_a}$ in the early Universe.
The initial conditions depend on the reheating process after inflation.
We assume that at the reheating temperature $T_\mathrm{reh}$, all SM particles are thermalized, while $n_{\Delta_a}$ and $n_{\bar{\Delta}_a}$ are both zero.

The maximum achievable reheating temperature $T_{\mathrm{reh}}^{\mathrm{max}}$ is bounded by the constraint on the tensor-to-scalar ratio $r$ from observations of the cosmic microwave background.
When the pivot scale $k_*$ exits the Hubble horizon, the inflationary potential can be expressed as \cite{Planck:2018jri,Baumann:2009ds}
\begin{equation}
  V_{\ast} = \frac{3 \pi^2}{2}
  A_{\mathrm{s}} r_{\ast} M^4_{\mathrm{Pl}},
\end{equation}
where $A_\mathrm{s}$ and $r_{\ast}$ are the amplitude of the primordial comoving curvature power spectrum and the tensor-to-scalar ratio at $k = k_*$.
At the pivot scale $k_* = 0.05~\si{Mpc}^{-1}$, the Planck data give $A_{\mathrm{s}} = 2.101^{+ 0.031}_{- 0.034} \times 10^{- 9}$ \cite{Planck:2018vyg}, while the BICEP/\textit{Keck} observations combined with Planck and WMAP data place a 95\% CL upper bound on the tensor-to-scalar ratio of $r_{0.05} < 0.036$ \cite{BICEP:2021xfz}.
These results constrain the energy scale of inflation to $V_{\ast}^{1 / 4} < 1.4 \times 10^{16}~\si{GeV}$.

If reheating is instantaneous, the radiation energy density at $T = T_\mathrm{reh}$ is $\rho_{\mathrm{r}} (T_{\mathrm{reh}}) = \rho_{\mathrm{end}}$, where $\rho_{\mathrm{end}}$ is the total energy density at the end of inflation, bounded as $\rho_{\mathrm{end}} \lesssim V_{\ast}$.
Using $\rho_{\mathrm{r}} (T_{\mathrm{reh}}) = \pi^2 g_{\ast} T_{\mathrm{reh}}^4/30$ with $g_{\ast} = 106.75$ accounting for the full SM contribution to the effective relativistic degrees of freedom, we derive an upper bound on the maximum achievable reheating temperature as
\begin{equation}
  T_{\mathrm{reh}}^{\mathrm{max}} \lesssim \left( \frac{30 V_{\ast}}{\pi^2
  g_{\ast}} \right)^{1 / 4} < 5.8 \times 10^{15}  ~\si{GeV}.
\end{equation}
If reheating is not instantaneous, the reheating temperature is expected to be lower.
In the following analysis, we consider two representative reheating temperatures: $T_\mathrm{reh} = 10^{15}~\si{GeV}$ and $10^{13}~\si{GeV}$.

In principle, the $CP$ phases in the Yukawa couplings \eqref{eq:Y10} and \eqref{eq:Y5} might generate an asymmetry between the number densities of $\Delta_a$ and $\bar{\Delta}_a$ particles.
Nonetheless, such an asymmetry is either difficult to produce or negligible in the scenario we consider.
We therefore safely assume $n_{\Delta_a} = n_{\bar{\Delta}_a}$.

The primary processes that alter the number density $n_{\Delta_a}$ are $\Delta_a \bar{\Delta}_b$ annihilation, $\Delta_a \Delta_c$ coannihilation ($c \neq a$), and $\Delta_a$ decays, together with their respective inverse processes.
The corresponding interaction rates are controlled by the thermally averaged cross sections $\langle \sigma_{\Delta_a\bar{\Delta}_b}v \rangle$ and $\langle\sigma_{\Delta_a\Delta_c}v\rangle$, as well as the thermally averaged decay width $\langle \Gamma_{\Delta_a} \rangle$.
Consequently, the evolution of $n_{\Delta_a}$ is governed by the Boltzmann equation \cite{Kolb:1990vq}
\begin{equation}
\frac{\mathrm{d}n_{\Delta_a}}{\mathrm{d}t} + 3 H n_{\Delta_a} = -  \left( \sum_{b} \langle\sigma_{\Delta_a\bar{\Delta}_b}v\rangle + \sum_{c\neq a} \langle\sigma_{\Delta_a\Delta_c}v\rangle \right) ( n_{\Delta_a}^2 - n_{\Delta_a,\mathrm{eq}}^2 ) - \left\langle \Gamma_{\Delta_a} \right\rangle \left( n_{\Delta_a} - n_{\Delta_{a},\mathrm{eq}} \right),
\end{equation}
where $H$ is the Hubble rate and $n_{\Delta_a,\mathrm{eq}}$ denotes the number density in thermal equilibrium.

Defining the total number density of scalar leptoquarks as
\begin{align}
    n = \sum_{a=1}^3 ( n_{\Delta_a} + n_{\bar{\Delta}_a} ) = 6n_{\Delta_a},
\end{align}
the corresponding Boltzmann equation takes the form
\begin{eqnarray}
\frac{\mathrm{d}n}{\mathrm{d}t} + 3Hn  & = & -\frac{1}{18} \left[ \sum_{a,b}\langle\sigma_{\Delta_a\bar{\Delta}_b}v\rangle + \sum_{a}\sum_{c\neq a}  \left(\langle\sigma_{\bar{\Delta}_a\bar{\Delta}_c}v\rangle + \langle\sigma_{\Delta_a\Delta_c}v\rangle \right) \right] ( n^2 - n_{\mathrm{eq}}^2 )
\nonumber\\
&& - \frac{1}{6} \sum_a \left( \langle \Gamma_{\Delta_a} \rangle + \langle \Gamma_{\bar{\Delta}_a} \rangle \right) ( n - n_{\mathrm{eq}} ).
\end{eqnarray}
We introduce the dimensionless variables $x \equiv m_\Delta / T$ and  $Y \equiv n/s$, where $s$ is the entropy density.
Here, $Y$ can be interpreted as the number density in a comoving volume.
In terms of these variables, the Boltzmann equation reads
\begin{eqnarray}
\frac{\mathrm{d}Y}{\mathrm{d}x} & = & -\frac{sx}{18H(m_\Delta)} \left[ \sum_{a,b}\langle\sigma_{\Delta_a\bar{\Delta}_b}v\rangle + \sum_{a}\sum_{c\neq a}  \left(\langle\sigma_{\bar{\Delta}_a\bar{\Delta}_c}v\rangle + \langle\sigma_{\Delta_a\Delta_c}v\rangle \right) \right] ( Y^2 - Y_{\mathrm{eq}}^2 )
\nonumber\\
&& - \frac{x}{6H(m_\Delta)} \sum_a \left( \langle \Gamma_{\Delta_a} \rangle + \langle \Gamma_{\bar{\Delta}_a} \rangle \right) ( Y-Y_{\mathrm{eq}} ),
\end{eqnarray}
where $H(m_\Delta)$ is the Hubble rate at $T = m_\Delta$.
This form is more convenient for numerical solution.

\begin{figure}[!t]
\centering
\subfigure{\includegraphics[scale=1]{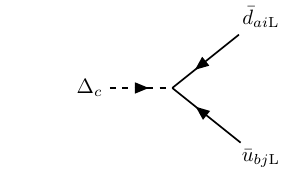}}
\subfigure{\includegraphics[scale=1]{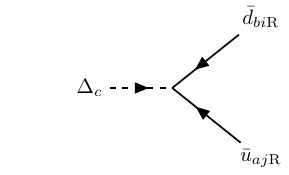}}
\subfigure{\includegraphics[scale=1]{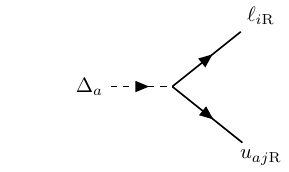}}
\subfigure{\includegraphics[scale=1]{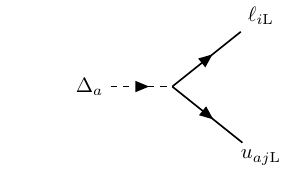}}
\subfigure{\includegraphics[scale=1]{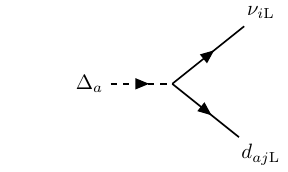}}
\caption{Feynman diagrams for two-body decays of the scalar leptoquarks $\Delta_a$.}
\label{fig:diag:Delta_2body_decay}
\end{figure}

Because of the Yukawa couplings in the Lagrangian \eqref{eq:L_Delta_Yuk}, the scalar leptoquarks can decay into SM fermions via two-body processes.
The related channels are $\Delta_c \to \bar{d}_{ai\mathrm{L}} \bar{u}_{bi\mathrm{L}}$, $\Delta_c \to \bar{d}_{bi\mathrm{R}} \bar{u}_{aj\mathrm{R}}$, $\Delta_a \to \ell_{i\mathrm{R}} u_{aj\mathrm{R}}$, $\Delta_a \to \ell_{i\mathrm{L}} u_{aj\mathrm{L}}$, and $\Delta_a \to \nu_{i\mathrm{L}} d_{aj\mathrm{L}}$, whose tree-level Feynman diagrams are illustrated in Fig.~\ref{fig:diag:Delta_2body_decay}.
Since the temperatures relevant to our calculation are far above the electroweak symmetry breaking scale, all SM fermions can be treated as massless.
The corresponding partial decay widths are
\begin{eqnarray}
  \Gamma_{\Delta_c \to \bar{d}_{ai \mathrm{L}}  \bar{u}_{b j \mathrm{L}}} & =
  & \frac{\varepsilon_{abc}^2 m_{\Delta}}{4 \pi}  | Y^{10}_{i j} + Y^{10}_{j
  i} |^2, \\
  \Gamma_{\Delta_c \to \bar{d}_{bi \mathrm{R}}  \bar{u}_{aj \mathrm{R}}} & = &
  \frac{\varepsilon_{abc}^2 m_{\Delta}}{32 \pi}  | Y^5_{i j} |^2, \\
  \Gamma_{\Delta_a \to \ell_{i \mathrm{R}} u_{aj \mathrm{R}}} & = &
  \frac{m_{\Delta}}{4 \pi} | Y^{10}_{i j} + Y^{10}_{j i} |^2, \\
  \Gamma_{\Delta_a \to \ell_{i \mathrm{L}} u_{aj \mathrm{L}}} & = & \Gamma_{\Delta_a \to \nu_{i \mathrm{L}} d_{aj \mathrm{L}}} = \frac{m_{\Delta}}{32
  \pi} | Y^5_{i j} |^2 . 
\end{eqnarray}
Note that the equality between $\Gamma_{\Delta_a \to \ell_{i \mathrm{L}} u_{aj \mathrm{L}}}$ and $\Gamma_{\Delta_a \to \nu_{i \mathrm{L}} d_{aj \mathrm{L}}}$ is guaranteed by the $\mathrm{SU}(2)_\mathrm{L}$ gauge symmetry.
The thermally averaged partial decay widths are then given  by \cite{Kolb:1979qa}
\begin{equation}
\langle \Gamma_i \rangle = \frac{\mathrm{K}_1(x)}{\mathrm{K}_2(x)}\, \Gamma_i,
\end{equation}
where $\mathrm{K}_1(x)$ and $\mathrm{K}_2(x)$ are modified Bessel functions of the second kind.

\begin{figure}[!t]
\centering
\subfigure{\includegraphics[scale=.92]{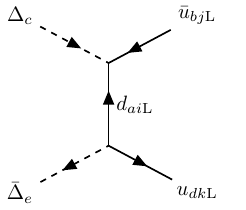}}
\subfigure{\includegraphics[scale=.92]{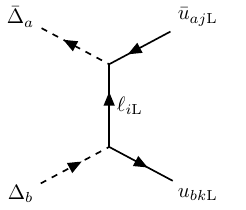}}
\subfigure{\includegraphics[scale=.92]{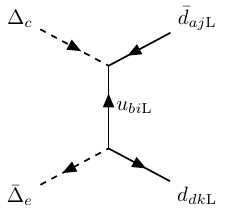}}
\subfigure{\includegraphics[scale=.92]{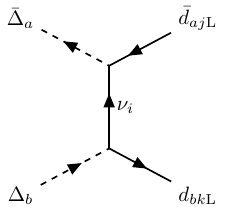}}
\subfigure{\includegraphics[scale=.92]{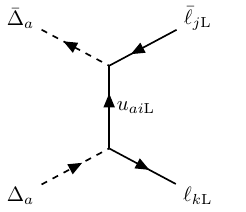}}
\subfigure{\includegraphics[scale=.92]{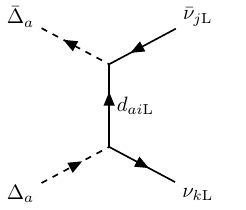}}
\subfigure{\includegraphics[scale=.92]{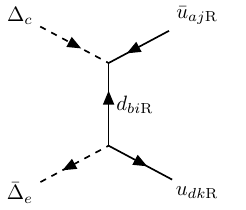}}
\subfigure{\includegraphics[scale=.92]{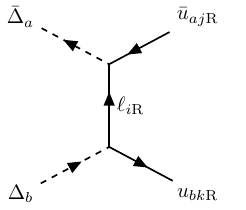}}
\subfigure{\includegraphics[scale=.92]{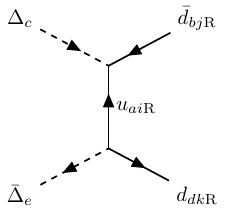}}
\subfigure{\includegraphics[scale=.92]{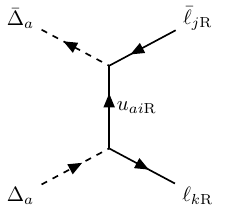}}
\caption{Feynman diagrams for $2\to 2$ annihilation processes of scalar leptoquarks through Yukawa couplings.}
\label{fig:diag:anni}
\end{figure}

\begin{figure}[!t]
\centering
\subfigure{\includegraphics[scale=.92]{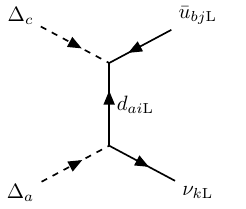}}
\subfigure{\includegraphics[scale=.92]{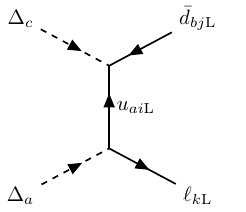}}
\subfigure{\includegraphics[scale=.92]{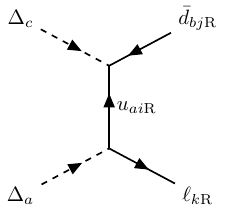}}
\caption{Feynman diagrams for $2\to 2$ coannihilation processes of scalar leptoquarks through Yukawa couplings.}
\label{fig:diag:coanni}
\end{figure}

Through the Yukawa couplings, the $2\to 2$ annihilation processes of scalar leptoquarks include $\Delta_a \bar{\Delta}_b \to u_{ci\mathrm{L}} \bar{u}_{dj\mathrm{L}}$, $\Delta_a \bar{\Delta}_b \to d_{ci\mathrm{L}} \bar{d}_{dj\mathrm{L}}$, $\Delta_a \bar{\Delta}_a \to \ell_{i\mathrm{L}} \bar{\ell}_{j\mathrm{L}}$, $\Delta_a \bar{\Delta}_a \to \nu_{i\mathrm{L}} \bar{\nu}_{j\mathrm{L}}$, $\Delta_a \bar{\Delta}_b \to u_{ci\mathrm{R}} \bar{u}_{dj\mathrm{R}}$, $\Delta_a \bar{\Delta}_b \to d_{ci\mathrm{R}} \bar{d}_{dj\mathrm{R}}$, and $\Delta_a \bar{\Delta}_a \to \ell_{i\mathrm{R}} \bar{\ell}_{j\mathrm{R}}$, while the $2\to 2$ coannihilation processes of scalar leptoquarks involve $\Delta_a \Delta_c \to \nu_{i\mathrm{L}} \bar{u}_{bj\mathrm{L}}$, $\Delta_a \Delta_c \to \ell_{i\mathrm{L}} \bar{d}_{bj\mathrm{L}}$, and $\Delta_a \Delta_c \to \ell_{i\mathrm{R}} \bar{d}_{bj\mathrm{R}}$.
The corresponding Feynman diagrams at tree level are shown in Figs.~\ref{fig:diag:anni} and \ref{fig:diag:coanni}.
For each process, we derive an analytic expression for its tree-level cross section $\sigma(s)$ as a function of the Mandelstam variable $s$.
The thermally averaged cross section is then obtained by numerically integrating the formula \cite{Gondolo:1990dk}
\begin{equation}
\left\langle\sigma v\right\rangle = \frac{x}{8 m_\Delta^5 \mathrm{K}_2^2(x)} \int_{4 m_\Delta^2}^{\infty} \sqrt{s}  (s-4 m_\Delta^2) \sigma(s) \,\mathrm{K}_1\!\left(\frac{x\sqrt{s}}{m_\Delta}\right) \mathrm{d} s .
\end{equation}

At high scales, the running of the Yukawa couplings \eqref{eq:Y10} and \eqref{eq:Y5} can be significant.
To reduce the theoretical uncertainties in the partial decay widths and cross sections, we therefore adopt the running Yukawa couplings of SM fermions, $y_f \equiv \sqrt{2} m_f/v$, together with the running CKM parameters, all evaluated in the $\overline{\mathrm{MS}}$ scheme at the scale $\mu = 10^{16}~\si{GeV}$ using two-loop renormalization group equations in the SM \cite{Antusch:2025fpm},
\begin{alignat}{3}
y_u(\mu) &= 2.87(7)\times 10^{-6},&~
y_d(\mu) &= 0.65(1)\times 10^{-5},&~
y_s(\mu) &= 1.29(2)\times 10^{-4},
\nonumber\\
y_c(\mu) &= 1.45(3)\times 10^{-3},&~
y_b(\mu) &= 0.606(6)\times 10^{-2},&~
y_t(\mu) &= 0.4454(48),
\nonumber\\
y_e(\mu) &= 2.6935(91)\times 10^{-6},&~
y_\mu(\mu) &= 5.6745(192)\times 10^{-4},&~
y_\tau(\mu) &= 0.9639(33)\times 10^{-2},
\nonumber\\
\theta_{12}^\mathrm{CKM}(\mu) &= 0.22708(82),&~
\theta_{23}^\mathrm{CKM}(\mu) &= 4.743(47)\times 10^{-2},&~
\theta_{13}^\mathrm{CKM}(\mu) &= 4.19(9)\times 10^{-3},
\nonumber\\
\delta^\mathrm{CKM}(\mu) &= 1.139(23).
\end{alignat}
The running effects on the PMNS matrix are neglected.

\begin{figure}[!t]
\centering
\subfigure[Total number density $n$ \label{fig:n_1e15}]{\includegraphics[width=0.48\textwidth]{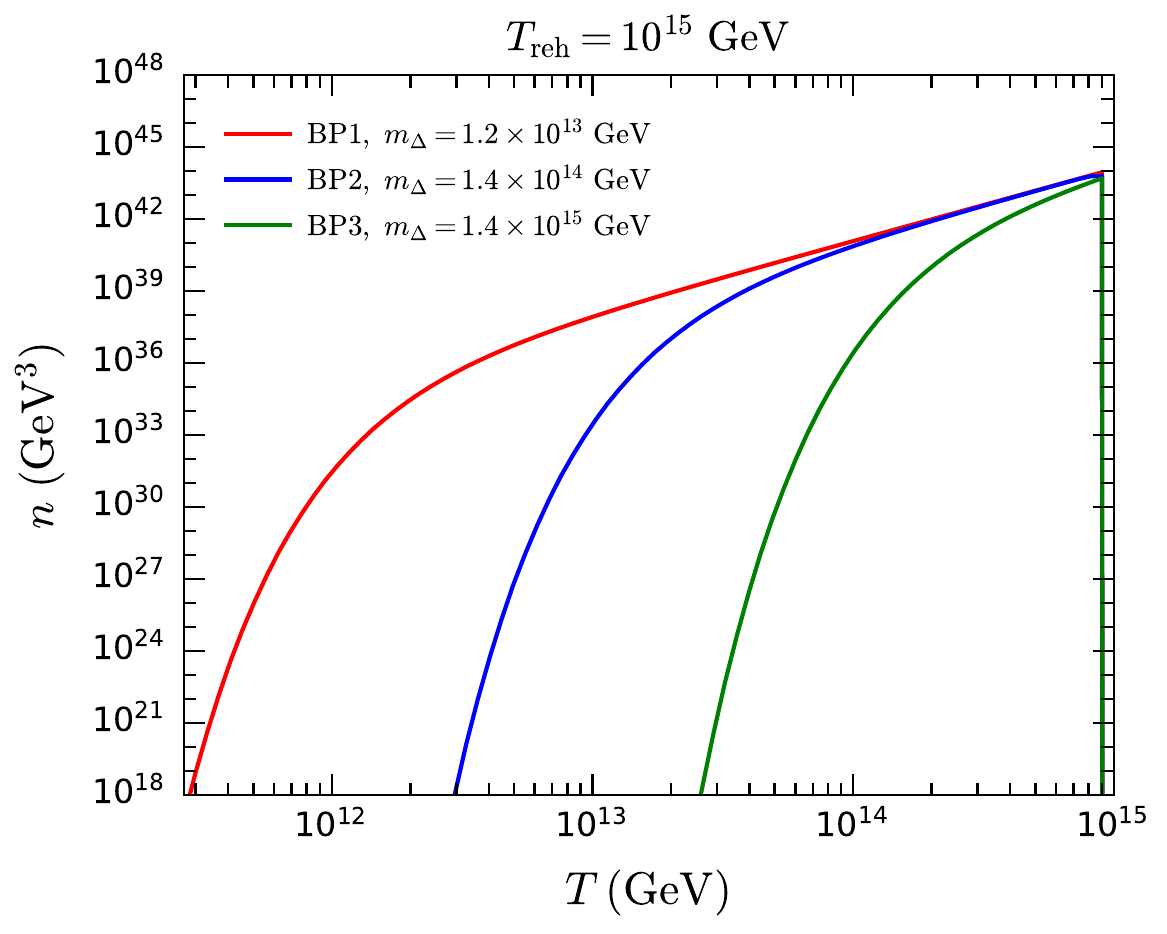}}
\hspace{0.02\textwidth}
\subfigure[$H,~\Gamma_\sigma,~\langle\Gamma\rangle$\label{fig:H_Gamma_1e15}]{\includegraphics[width=0.48\textwidth]{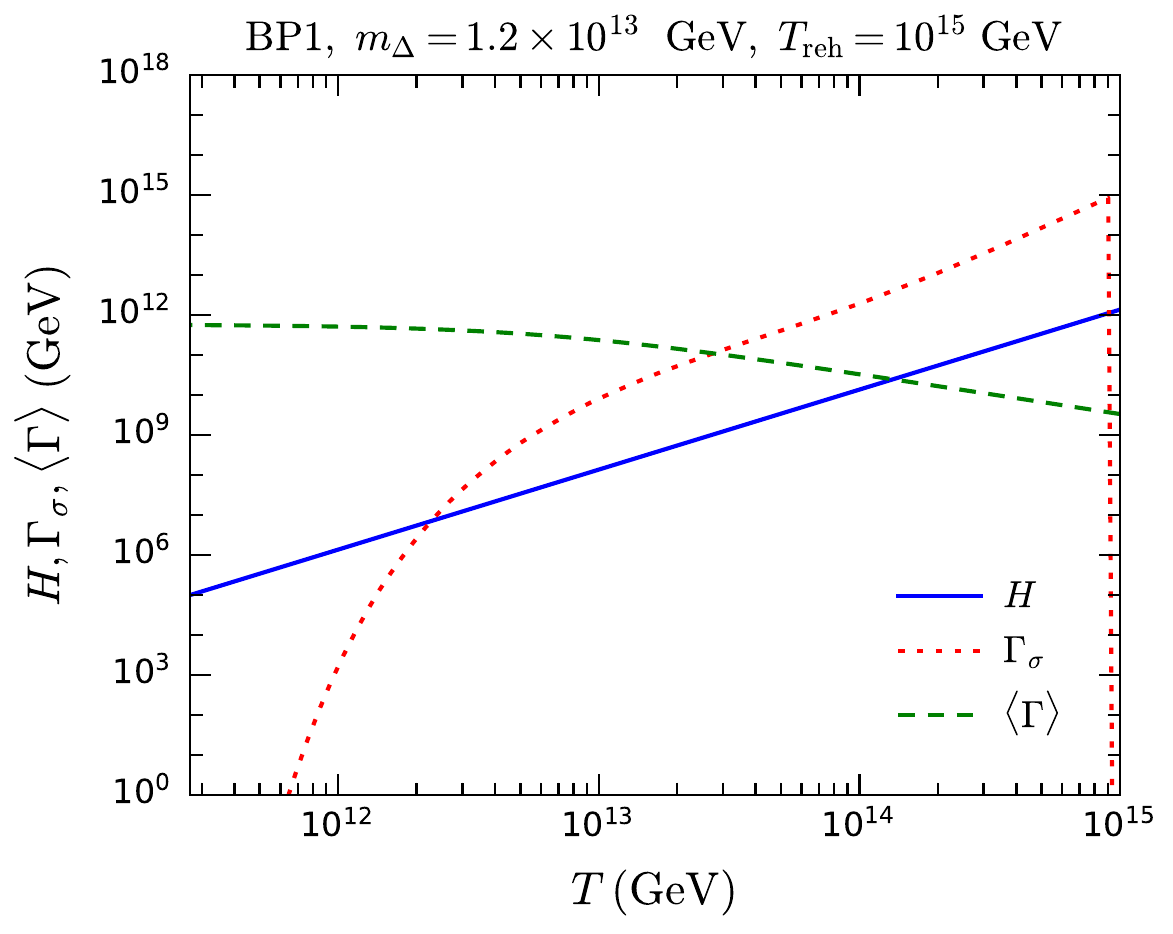}}
\caption{Evolution of (a) the total number density of scalar leptoquarks for three BPs, and (b) the Hubble rate $H$, the total annihilation and coannihilation rate $\Gamma_\sigma$, and the total decay rate $\left\langle\Gamma\right\rangle$ for BP1, assuming $T_\mathrm{reh}= 10^{15}~\mathrm{GeV}$.}
\label{fig:n_H_Gamma_1e15}
\end{figure}

We numerically solve the Boltzmann equation to obtain the total number densities of scalar leptoquarks, $n$, as functions of $T$ for the three BPs.
The results for $T_\mathrm{reh}=10^{15}\mathrm{~GeV}$ are shown in Fig.~\ref{fig:n_1e15}.
We find that the number densities attain thermal equilibrium shortly after reheating and thereafter closely track their equilibrium values.
Consequently, $n$ drops rapidly once $T \ll m_\Delta$.
This behavior indicates that thermal equilibrium is maintained by sufficiently strong interaction rates of the scalar leptoquarks, as can be understood by comparing the interaction rates with the Hubble rate $H$.

The total annihilation and coannihilation rate of scalar leptoquarks is defined as
\begin{equation}
\Gamma_\sigma = n \left[
\sum_{a,b}\langle\sigma_{\Delta_a\bar{\Delta}_b}v\rangle + \sum_{a}\sum_{c\neq a}  \left(\langle\sigma_{\bar{\Delta}_a\bar{\Delta}_c}v\rangle + \langle\sigma_{\Delta_a\Delta_c}v\rangle \right)\right],
\end{equation}
while their total decay rate is given by
\begin{equation}
\langle\Gamma\rangle = \sum_a \left( \langle \Gamma_{\Delta_a} \rangle + \langle \Gamma_{\bar{\Delta}_a} \rangle \right).
\end{equation}
For BP1, the evolution of these rates, together with the Hubble rate, is demonstrated in Fig.~\ref{fig:H_Gamma_1e15}.
We find that $\Gamma_\sigma > H$ for $2\times 10^{12}~\mathrm{GeV} \lesssim T < T_\mathrm{reh}$ and $\langle\Gamma\rangle > H$ for $T\lesssim 10^{14}\mathrm{~GeV}$.
Thus, for $T < T_\mathrm{reh}$, the total interaction rate remains sufficiently high to keep the scalar leptoquarks in thermal equilibrium.
The situation for BP2 and BP3 is similar to that of BP1.

\begin{figure}[!t]
\includegraphics[width=0.48\textwidth]{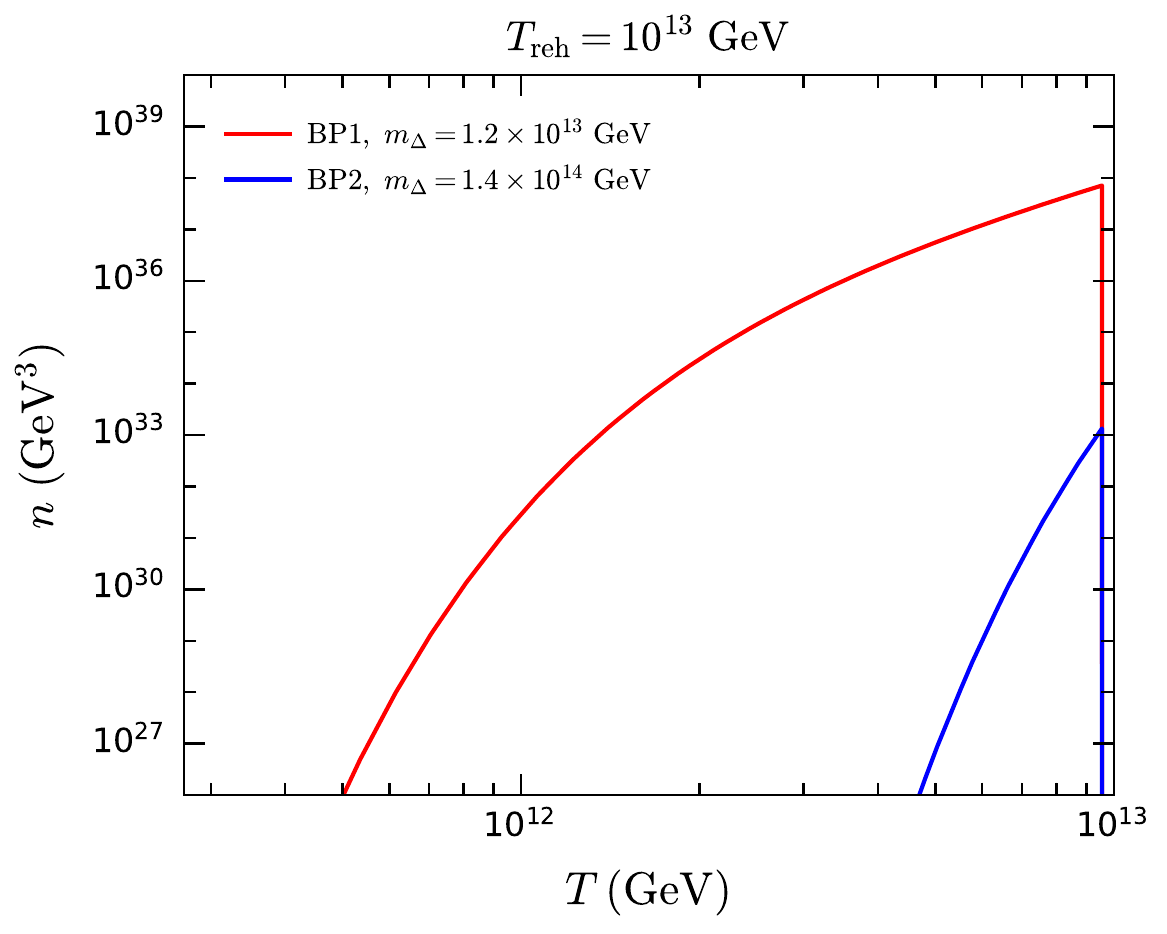}
\caption{Evolution of the total number density of scalar leptoquarks for BP1 and BP2, assuming $T_{\mathrm{reh}} = 10^{13}~\mathrm{GeV}$.}
\label{fig:n_1e13}
\end{figure}

Furthermore, for $T_{\mathrm{reh}} = 10^{13}~\mathrm{GeV}$, the total number densities of scalar leptoquarks for BP1 and BP2, as functions of the temperature $T$, are presented in Fig.~\ref{fig:n_1e13}.
Shortly after reheating, they reach thermal equilibrium and maintain it thereafter.
For BP3 with $m_\Delta = \num{1.42e15}~\si{GeV}$, the number density is negligible and is therefore not shown, since its equilibrium value is already strongly suppressed by the Boltzmann factor $\exp(-m_\Delta/T)$ after reheating.

However, the above analysis is not fully complete.
Since the scalar leptoquarks $\Delta_a$ form an $\mathrm{SU}(3)_\mathrm{C}$ triplet with weak hypercharge $Y = -1/3$, the gauge couplings also lead to $\Delta_a \bar{\Delta}_b$ annihilation into SM fermions, gluons, and $\mathrm{U}(1)_\mathrm{Y}$ gauge bosons.
Therefore, the actual annihilation and coannihilation rate $\Gamma_\sigma$ would exceed the values obtained above, which include only the contributions from Yukawa couplings.
Nevertheless, this does not alter the key conclusion that the scalar leptoquark number density tracks its equilibrium value shortly after reheating.
Thus, the number density obtained above remains safely applicable.

\section{Graviton Bremsstrahlung and Gravitational Waves}
\label{sub:GW}

In this section, we estimate the stochastic GW spectrum generated by graviton bremsstrahlung in scalar leptoquark decays.
This signal originates from the gravitational interaction between the graviton and the matter fields.
Based on Einstein's general relativity, the gravitational action is given by
\begin{equation}
  S = \int \mathrm{d}^4 x \, \sqrt{- g} \left( \frac{M_{\mathrm{Pl}}^2}{2}\, R
  +\mathcal{L} \right),
\end{equation}
where $R$ is the Ricci scalar and $g = \det(g_{\mu\nu})$ denotes the determinant of the metric tensor $g_{\mu\nu}$.
The reduced Planck mass $M_{\mathrm{Pl}}$ is related to the Newtonian gravitational constant $G_\mathrm{N}$ by $M_{\mathrm{Pl}} = (8 \pi G_{\mathrm{N}})^{-1/2} = \num{2.435e+18}~\si{GeV}$.
The Lagrangian $\mathcal{L}$ encompasses all fields in the $\mathrm{SU}(5)$ GUT.

When the gravitational field is weak, one can work within the framework of linearized gravity.
In this regime, the metric tensor can be decomposed as
\begin{equation}\label{eq:g_h}
g_{\mu\nu} = \eta_{\mu\nu} + \kappa h_{\mu\nu},
\end{equation}
where $\eta_{\mu\nu}$ is the Minkowski metric and $\kappa \equiv 2/M_\mathrm{Pl}$ is the gravitational coupling constant.
The rank-2 symmetric tensor field $h_{\mu\nu}$ represents a small perturbation around the flat background, and, upon quantization, corresponds to the massless spin-2 graviton field.
With the expansion in Eq.~\eqref{eq:g_h}, any curved-space geometric quantity can be expressed as a power series in $h_{\mu\nu}$.
One can therefore construct the gravitational interaction Lagrangian to any order in $\kappa$ \cite{Choi:1994ax}.
The resulting Feynman rules for the graviton couplings to spin-0, spin-1/2, and spin-1 particles at $\mathcal{O}(\kappa)$ are given in Ref.~\cite{Barman:2023ymn}.

\begin{figure}[!t]
\centering
\includegraphics[scale=1]{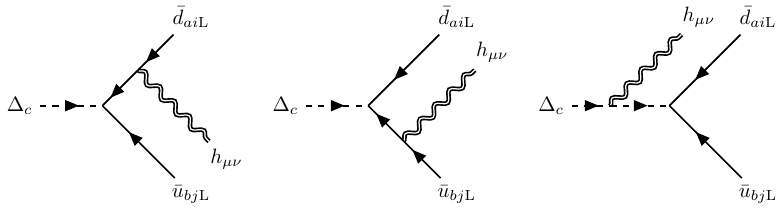}
\caption{Feynman diagrams for the graviton bremsstrahlung process $\Delta_c \to \bar{d}_{ai\mathrm{L}} \bar{u}_{bj\mathrm{L}} h$.}			\label{fig:feynman_figure_Delta2h_bar_dL_bar_uL}
\end{figure}

Within this framework, we can compute the differential decay rate for graviton bremsstrahlung accompanying the decays of scalar leptoquarks.
As an explicit example, the lowest‑order Feynman diagrams for the process $\Delta_c \to \bar{d}_{ai\mathrm{L}} \bar{u}_{bj\mathrm{L}} h$, where $h$ denotes the graviton, are shown in Fig.~\ref{fig:feynman_figure_Delta2h_bar_dL_bar_uL}.
Similar diagrams can be obtained by attaching a graviton line to the other two-body decay diagrams in Fig.~\ref{fig:diag:Delta_2body_decay}.
The differential decay rates for $\Delta_c \to \bar{d}_{ai\mathrm{L}} \bar{u}_{bi\mathrm{L}} h$, $\Delta_c \to \bar{d}_{bi\mathrm{R}} \bar{u}_{aj\mathrm{R}} h$, $\Delta_a \to \ell_{i\mathrm{R}} u_{aj\mathrm{R}} h$, $\Delta_a \to \ell_{i\mathrm{L}} u_{aj\mathrm{L}} h$, and $\Delta_a \to \nu_{i\mathrm{L}} d_{aj\mathrm{L}} h$ are given by \cite{Nakayama:2018ptw,Barman:2023ymn}
\begin{eqnarray}
\frac{\mathrm{d} \Gamma_{\Delta_c \to \bar{d}_{ai \mathrm{L}}  \bar{u}_{bj
  \mathrm{L}} h}}{\mathrm{d} E} & = & \frac{\varepsilon_{a b c}^2 | Y_{i j}^{10} +
  Y_{j i}^{10} |^2 m_{\Delta}^2}{96 \pi^3 M_{\mathrm{Pl}}^2} \, f \left(
  \frac{E}{m_{\Delta}} \right),
\\
\frac{\mathrm{d} \Gamma_{\Delta_c \to \bar{d}_{bi \mathrm{R}}  \bar{u}_{aj
  \mathrm{R}} h}}{\mathrm{d} E} & = & \frac{\varepsilon_{a b c}^2 | Y_{i j}^5 |^2
  m_{\Delta}^2}{768 \pi^3 M_{\mathrm{Pl}}^2} \, f \left( \frac{E}{m_{\Delta}}
  \right),
\\
\frac{\mathrm{d} \Gamma_{\Delta_a \to \ell_{i \mathrm{R}} u_{aj \mathrm{R}}
  h}}{\mathrm{d} E} & = & \frac{| Y_{i j}^{10} + Y_{j i}^{10} |^2 m_{\Delta}^2}{96
  \pi^3 M_{\mathrm{Pl}}^2} \, f \left( \frac{E}{m_{\Delta}} \right),
\\
  \frac{\mathrm{d} \Gamma_{\Delta_a \to \ell_{i \mathrm{L}} u_{aj \mathrm{L}}
  h}}{\mathrm{d} E} & = & \frac{\mathrm{d} \Gamma_{\Delta_a \to \nu_{i \mathrm{L}}
  d_{aj \mathrm{L}} h}}{\mathrm{d} E} = \frac{| Y_{i j}^5 |^2 m_{\Delta}^2}{768
  \pi^3 M_{\mathrm{Pl}}^2} \, f \left( \frac{E}{m_{\Delta}} \right),
\end{eqnarray}
where $E$ is the energy of the bremsstrahlung graviton, and
\begin{equation}
f (x) \equiv \frac{1 - 2 x}{x} (1 - 2 x + 2 x^2).
\end{equation}
Note that these differential decay rates scale as $m_\Delta^2/M_\mathrm{Pl}^2$.
Therefore, the graviton bremsstrahlung processes are enhanced for heavier scalar leptoquarks.

The incoherent superposition of bremsstrahlung gravitons emitted from scalar leptoquark decays forms an SGWB.
Because of the cosmological redshift, the graviton energy $E(t)$ at cosmic time $t$ is related to its present-day value $E_0$ by $E_0 = E (t) a (t)$, where $a(t)$ is the cosmological scale factor normalized to $a(t_0) = 1$  at the present time $t_0$.
Denoting the present energy density of the SGWB as $\rho_\mathrm{GW}$, we have \cite{Hu:2024awd}
\begin{align}
\frac{\mathrm{d}\rho_{\mathrm{GW}}}{\mathrm{d}E_0} = E_0 \int \mathrm{d}t \, n(t)\, a^2(t)\, \frac{\mathrm{d}\Gamma_h}{\mathrm{d}E},
\end{align}
where $n(t)$ is the total number density of scalar leptoquarks obtained in the previous section, and ${\mathrm{d}\Gamma_h}/{\mathrm{d}E}$ is the total differential decay rate for graviton bremsstrahlung in scalar leptoquark decays.

The scale factor $a(t)$ as a function of cosmic time can be obtained by solving
\begin{equation}
\frac{\mathrm{d}a(t)}{\mathrm{d}t} = a(t) H(t),
\end{equation}
where the Hubble rate $H(t)$ in the $\Lambda \mathrm{CDM}$ cosmological model is given by \cite{Binetruy:2012ze}
\begin{equation}
H(t) = H_0 \sqrt{\Omega_\mathrm{r} \mathcal{G}(z)a^{-4}(t) + \Omega_\mathrm{m} a^{-3}(t)+\Omega_{\Lambda}}.
\end{equation}
Here, $H_0$ is the Hubble constant, and $z = a^{-1}(t) - 1$ is the redshift.
The parameters $\Omega_\mathrm{r} = 1.68\times 5.38(15)\times 10^{-5}$, $\Omega_\mathrm{m} = 0.315(7)$, and $\Omega_\Lambda = 0.685(7)$  \cite{ParticleDataGroup:2024cfk} denote the present energy density fractions of radiation, matter, and dark energy relative to the critical density $\rho_\mathrm{c} \equiv 3H_0^2/(8\pi G_\mathrm{N})$.
The function
\begin{equation}
\mathcal{G}(z) = \frac{g_\star(z)g_{\star s}^{4/3}(0)}{g_\star(0)g_{\star s}^{4/3}(z)}
\end{equation}
accounts for the change in the number of radiation degrees of freedom between redshift $z$ and the present, with $g_\star$ and $g_{\star s}$ the effective relativistic degrees of freedom for energy density and entropy density.
It is well-approximated by the piecewise form
\begin{equation}
\mathcal{G}(z) =
\left\{\begin{array}{ll}
1, &\quad z<10^9 \\
0.83, &\quad 10^9<z<2\times 10^{12},\\
0.39, &\quad z>2\times 10^{12}
\end{array}\right.
\end{equation}
where the transitions at $z = 10^9$ and $z = 2\times 10^{12}$ roughly correspond to the epochs of electron-positron annihilation and the QCD phase transition, respectively.

The relation between the present-day graviton energy $E_0$ and the GW frequency $f$ is $E_0 = 2 \pi f$.
It is conventional to express the SGWB frequency spectrum through the dimensionless quantity $\Omega_{\mathrm{GW}}(f)h^2$, with
\begin{equation}
  \Omega_{\mathrm{GW}} (f) \equiv \frac{f}{\rho_{\mathrm{c}}} \frac{\mathrm{d}
  \rho_{\mathrm{GW}}}{\mathrm{d} f} = \frac{E_0}{\rho_{\mathrm{c}}} \frac{\mathrm{d}
  \rho_{\mathrm{GW}}}{\mathrm{d} E_0}.
\end{equation}
Here, $h = 0.674(5)$ \cite{ParticleDataGroup:2024cfk} is the Hubble constant $H_0$ in units of $100~\si{km~s^{-1}~Mpc^{-1}}$.

\begin{figure}[!t]
\centering
\subfigure[$T_\mathrm{reh} = 10^{15}~\mathrm{GeV}$ \label{fig:dOmega_dlnE0_avg_1e15}]{\includegraphics[width=0.48\textwidth]{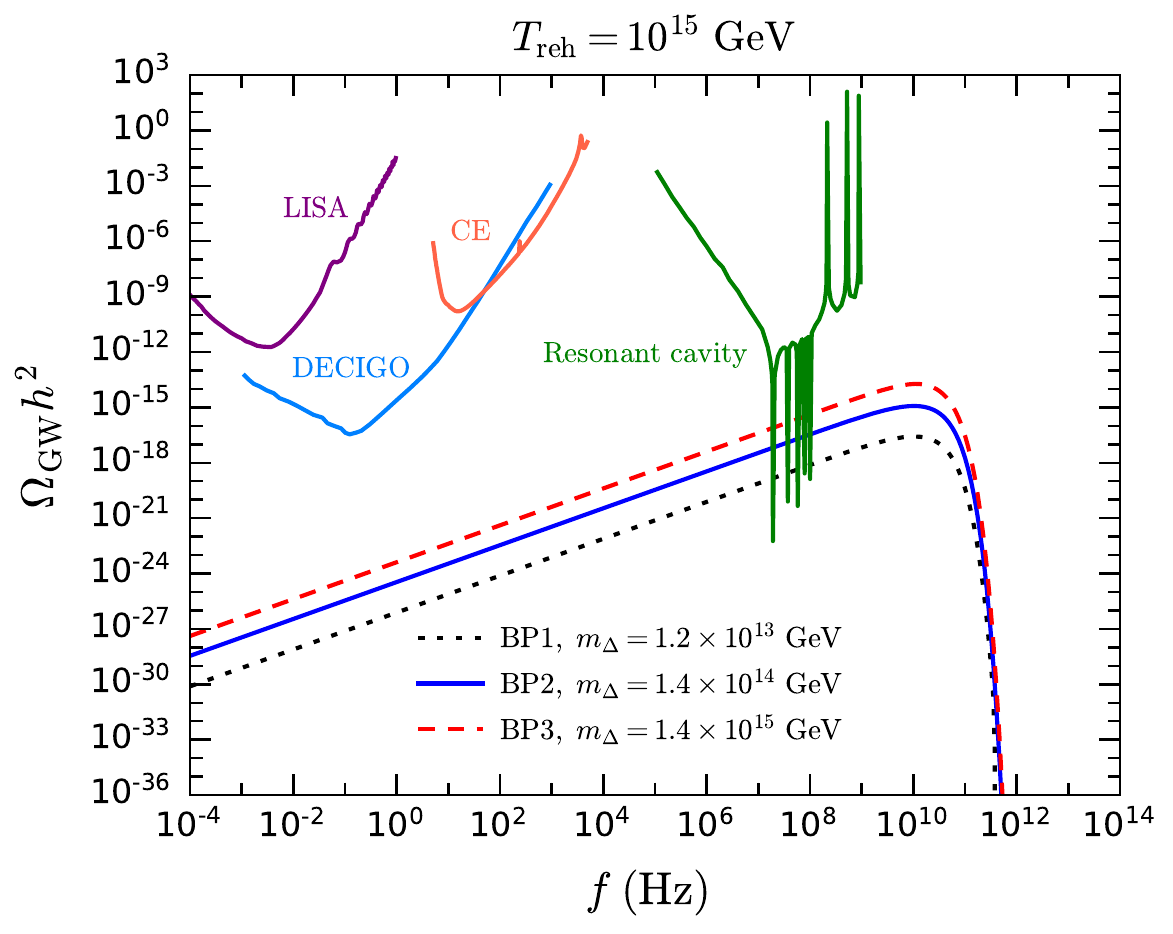}}
\hspace{0.02\textwidth}
\subfigure[$T_\mathrm{reh} = 10^{13}~\mathrm{GeV}$ \label{fig:dOmega_dlnE0_avg_1e13}]{\includegraphics[width=0.48\textwidth]{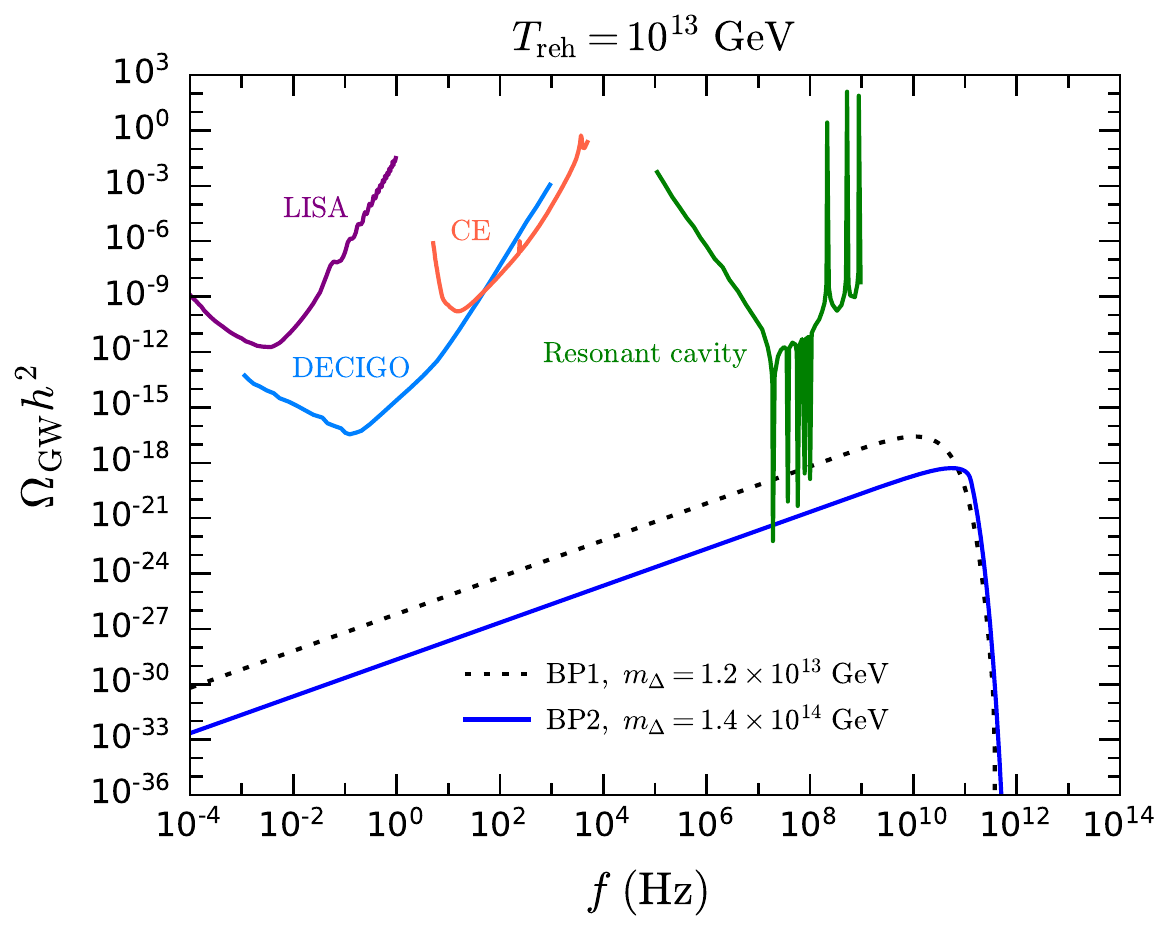}}
\caption{SGWB spectra originating from graviton bremsstrahlung in scalar leptoquark decays for the BPs, assuming $T_\mathrm{reh}=10^{15}~\mathrm{GeV}$ (a) and $10^{13}~\mathrm{GeV}$ (b).
Sensitivity curves of CE \cite{LIGOScientific:2016wof}, LISA \cite{LISA:2017pwj}, DECIGO \cite{Kawamura:2023flo}, and resonant cavity experiments \cite{Herman:2022fau} are also displayed.}
\label{fig:dOmega_dlnE0_avg}
\end{figure}

Based on the above formalism, we calculate the SGWB spectra for the BPs, as illustrated in Figs.~\ref{fig:dOmega_dlnE0_avg_1e15} and \ref{fig:dOmega_dlnE0_avg_1e13}, corresponding to $T_\mathrm{reh}=10^{15}~\mathrm{GeV}$ and $10^{13}~\mathrm{GeV}$, respectively.
For $T_\mathrm{reh}=10^{15}~\mathrm{GeV}$, the spectra of the three BPs all peak at around $10~\si{GHz}$, with cutoff frequencies of $\sim 500~\si{GHz}$.
As mentioned, the graviton bremsstrahlung rate is enhanced for heavier scalar leptoquarks, which is why the spectrum for BP3 has the largest amplitude.
For the lower reheating temperature $T_\mathrm{reh}=10^{13}~\mathrm{GeV}$, the scalar leptoquark number densities are suppressed.
The suppression is mild for BP1 but significant for BP2.
Consequently, the SGWB spectrum for BP1 is only slightly lower than that in the $T_\mathrm{reh}=10^{15}~\mathrm{GeV}$ case, whereas the spectrum for BP2 is reduced by three orders of magnitude.
In addition, the SGWB spectrum for BP3 is entirely negligible and is therefore not shown.

For comparison, we also plot the sensitivity curves of future GW experiments, including Cosmic Explorer (CE) \cite{LIGOScientific:2016wof}, Laser Interferometer Space Antenna (LISA) \cite{LISA:2017pwj}, Deci-hertz Interferometer Gravitational Wave Observatory (DECIGO) \cite{Kawamura:2023flo}, and resonant cavity experiments \cite{Herman:2022fau}.
A resonant cavity detector \cite{Herman:2020wao,Berlin:2021txa,Herman:2022fau,Berlin:2023grv,Gatti:2024mde,Cai:2025fpe} can probe SGWB signals in the MHz to GHz band through the inverse Gertsenshtein effect \cite{Gertsenshtein}, in which incoming GWs passing through a static magnetic field induce electromagnetic excitations inside a cavity.
In particular, the resonant modes around $20\text{--}100~\si{MHz}$ in such experiments are promising for detecting the SGWB signals from all three BPs if $T_\mathrm{reh} = 10^{15}~\mathrm{GeV}$, and from BP1 if $T_\mathrm{reh} = 10^{13}~\mathrm{GeV}$, as shown in Fig.~\ref{fig:dOmega_dlnE0_avg}.

\section{Conclusions}
\label{sub:Conclusions}

In this work, we have studied the SGWB generated by graviton bremsstrahlung in decays of scalar leptoquarks.
Scalar leptoquarks naturally arise in the Higgs sectors of GUTs.
Their interactions violate both lepton and baryon numbers, mediating proton decay, which pushes their masses well beyond the electroweak scale into the superheavy regime.
Because of quantum gravitational processes, decays of these superheavy particles in the early Universe are inevitably accompanied by graviton bremsstrahlung.
The resulting graviton emissions produce an SGWB that carries information from early epochs to present-day observatories.

As an explicit example, we considered the scalar leptoquarks $\Delta_a$ contained in the $\mathbf{5}_\mathrm{H}$ Higgs multiplet of an minimal, renormalizable $\mathrm{SU}(5)$ GUT, which is capable of reproducing viable fermion masses and mixing patterns.
To obtain quantitative predictions, we performed a random scan over the parameter space and impose current experimental limits on proton decay.
We found that the $p\to K^+ \bar{\nu}$ channel gives the strongest bound, excluding parameter points with $m_\Delta \lesssim 10^{13}~\si{GeV}$.
Among the parameter points that satisfy the proton decay bounds, we selected three BPs with scalar leptoquark masses $m_\Delta = \num{1.15e13}$, $\num{1.41e14}$, and $\num{1.42e15}~\si{GeV}$ for further analysis.

We then computed the thermal evolution of the scalar leptoquark number density by solving the Boltzmann equation, including annihilation, coannihilation, and decay processes, assuming two representative reheating temperatures, $T_\mathrm{reh} = 10^{15}~\si{GeV}$ and $10^{13}~\si{GeV}$.
We found that shortly after reheating, the resulting number density closely tracks its equilibrium value, because the total interaction rate is larger than the Hubble rate throughout the regime of interest.
The SGWB spectra originated from graviton bremsstrahlung in scalar leptoquark decays are further derived.
For $T_\mathrm{reh} = 10^{15}~\si{GeV}$, all three BPs yield significant SGWB spectra in the high-frequency range, peaking around $10~\si{GHz}$.
For $T_\mathrm{reh} = 10^{13}~\si{GeV}$, however, the scalar leptoquark number density for BP2 is suppressed, and the SGWB spectrum is reduced by three orders of magnitude, while the SGWB spectrum for BP3 is negligible.
Importantly, a portion of these spectra falls within the sensitivity band of proposed high-frequency GW detectors that employ resonant cavity techniques operating in the MHz to GHz range.
In particular, the resonant modes around $20\text{--}100~\si{MHz}$ offer promising detection prospects for the signals from all three BPs for $T_\mathrm{reh} = 10^{15}~\mathrm{GeV}$, and from BP1 for $T_\mathrm{reh} = 10^{13}~\mathrm{GeV}$.

Our findings demonstrate that gravitational waves from graviton bremsstrahlung provide a novel and complementary probe of superheavy particles far beyond the reach of terrestrial colliders.
A detection of such a high-frequency SGWB might not only reveal the existence of scalar leptoquarks but also provide valuable insights into the physics of grand unification and the quantum nature of gravity.

\begin{acknowledgments}

The authors thank Jiaming Zheng and Ye-Ling Zhou for helpful discussions.
This work is supported by the National Natural Science Foundation of China under Grant No. 12575115.

\end{acknowledgments}

\bibliographystyle{utphys}
\bibliography{ref}

\end{document}